\documentclass[a4paper,11pt]{article}
\pdfoutput = 1
\usepackage{jheppub}
\usepackage{amsmath,amssymb,amscd,braket,amsfonts}
\usepackage{color}
\usepackage{graphicx}
\usepackage{slashed}
\usepackage{amsthm}
\usepackage{enumerate}
\usepackage[mathscr]{euscript}
\usepackage{fancyvrb}
\usepackage{nicematrix}
\VerbatimFootnotes

\def\p{\partial}
\def\({\left(}
\def\){\right)}
\newcommand{\bea}{\begin{eqnarray}}
\newcommand{\eea}{\end{eqnarray}}
\newcommand{\be}{\begin{equation}}
\newcommand{\ee}{\end{equation}}
\newcommand{\ba}{\begin{align}}
\newcommand{\ea}{\end{align}}
\newcommand{\im}{\mathfrak{Im}}
\newcommand{\re}{\mathfrak{Re}}

\usepackage{amstext} 
\usepackage{array}   
\newcolumntype{L}{>{$}l<{$}} 
\newcolumntype{C}{>{$}c<{$}} 

\title{Transient dynamics of quasinormal mode sums}
\date{2024}

\author{Javier Carballo}
\emailAdd{j.carballo@soton.ac.uk}

\author{and Benjamin Withers}
\emailAdd{b.s.withers@soton.ac.uk}

\affiliation{Mathematical Sciences and STAG Research Centre, University of Southampton, Highfield, Southampton SO17 1BJ, UK}

\abstract{
Quasinormal modes of spacetimes with event horizons are typically governed by a non-normal operator.
This gives rise to spectral instabilities, a topic of recent interest in the black hole pseudospectrum programme.
In this work we show that non-normality leads to the existence of arbitrarily long-lived sums of short-lived quasinormal modes, corresponding to localising packets of energy near the future horizon. 
There exist sums of $M$ quasinormal modes whose lifetimes scale as $\log{M}$.
This transient behaviour results from large cancellations between non-orthogonal quasinormal modes.
We provide simple closed-form examples for a massive scalar field in the static patch of dS$_{d+1}$ and the BTZ black hole.
We also provide numerical examples for scalar perturbations of Schwarzschild-AdS$_{d+1}$, and gravitational perturbations of Schwarzschild in asymptotically flat spacetime, using hyperboloidal foliations.
The existence of these perturbations is linked to certain properties of black hole pseudospectra.
We comment on implications for thermalisation times in holographic plasmas.
}

\begin{document}
\maketitle
\flushbottom 

\section{Introduction}

Normal operators are defined by having a complete and orthogonal set of eigenfunctions with respect to a given inner product \cite{Kato:1966:PTL}. In the context of linear dynamics, the maximum response that can be developed from initial data depends strongly on whether the Hamiltonian operator is normal or non-normal. In the normal case, the maximum response is given by the slowest-decaying eigenfunction, whereas in the non-normal case the maximum response is generically unbounded.

In this work we ask the question of what is the maximum response that can be developed in linear perturbations of spacetimes with event horizons. We show that non-normality in these systems leads to novel transient phenomena which can be arbitrarily long-lived, meriting further investigation as observable phenomena in cosmology, gravitational wave physics, analogue gravity, and strongly-coupled many-body systems.

The Orr-Sommerfeld equation in hydrodynamics \cite{Orr1907,sommerfeld1909beitrag} is a paradigmatic example of the effects of non-normality, historically relevant in the study of the transition from laminar to turbulent flows. It governs linearised perturbations to the Navier-Stokes equations describing viscous parallel shear flows. In particular, perturbed plane Poiseuille flow was found to become linearly unstable at a critical Reynolds number $Re_c=5772.22$ \cite{Orszag1971AccurateSO}, when one of the eigenvalues of the Orr-Sommerfeld operator crosses into the unstable half of the complex plane. However, experiments showed that transition to turbulence happens at much lower values of $Re$, discarding this instability as the precursor to turbulence (see \cite{Trefethen1993} for an overview). A broader study, not restricted to the dynamical stability of the eigenvalues, revealed that the energy of the disturbance exhibits transient growth\footnote{A note on nomenclature. In this work we consider systems which decay at asymptotically late times with rates determined by eigenvalues of the Hamiltonian. We refer to behaviour preceding this `modal' regime as `transient'.} at Reynolds numbers well below the critical value \cite{Reddy93, Gustavsson_1991, Henningson_Lundbladh_Johansson_1993, ButlerFarrell, Reddy_Henningson_1993, Trefethen1993}, hence being a better indicator of the onset of turbulent dynamics. This transient growth is inherently associated to the non-orthogonality of the Orr-Sommerfeld eigenfunctions, as large cancellations can happen between the coefficients for the eigenfunction expansion of initial data that need not survive at later times. 

Recently, there has been great interest in the non-normality of operators governing black hole linear perturbation theory. Much focus has been placed on the consequences for spectral stability of black hole quasinormal modes (QNMs), regarding the distance their eigenvalues move in the complex plane under small static perturbations of the QNM radial operator. Spectral stability of QNM frequencies is not a recent issue, dating back to the seminal works \cite{Nollert:1996rf,Nollert:1998ys}. Nonetheless, it had never been addressed using pseudospectrum methods prior to \cite{Jaramillo:2020tuu}, where the Pöschl-Teller potential (governing perturbations of the dS$_{2}$ static patch) and Schwarzschild black hole were considered, paving the way for an intensive QNM pseudospectrum programme \cite{Jaramillo:2021tmt, Destounis:2021lum, Gasperin:2021kfv, Cheung:2021bol, Boyanov:2022ark, Sarkar:2023rhp, Arean:2023ejh, Courty:2023rxk, Cownden:2023dam, Destounis:2023nmb, Boyanov:2023qqf, Cao:2024oud}.\footnote{See also \cite{Rosato:2024arw, Oshita:2024fzf} for recent related work studying the effects of environmental perturbations on greybody factors.} Broadly speaking, the qualitative features of pseudospectra indicative of spectral instability were seen across a variety of spacetimes, extending the results of \cite{Jaramillo:2020tuu} to asymptotically AdS black holes \cite{Arean:2023ejh,Cownden:2023dam,Boyanov:2023qqf}, asymptotically dS black holes \cite{Sarkar:2023rhp,Destounis:2023nmb}, and other contexts \cite{Destounis:2021lum, Boyanov:2022ark}.\footnote{In this paper we also extend these findings to the static patch of dS$_{d+1}$ with $d>1$.} See also \cite{Kokkotas:1999bd, Berti:2009kk} for reviews of more general aspects of QNMs.

While the qualitative aspects of black hole pseudospectra are a striking embodiment of the effects of non-normality of black hole perturbations, the physical significance of such features are so far less evident. In particular, pseudospectra appear to be strongly dependent on an arbitrary choice of norm, and a generic choice is not numerically convergent \cite{Boyanov:2023qqf}. Furthermore, one may question the extent to which a destabilising static separable perturbation can exist in a physical context (see \cite{Cardoso:2024mrw} for a recent discussion and some examples).

We study consequences of non-normality of QNMs on what we believe to be more direct physical grounds. Namely, following the Orr-Sommerfeld example and importing many of the mathematical tools utilised there \cite{Farrell88, Reddy93, Reddy_Henningson_1993, TrefethenEmbree2005, Trefethen1993, SchmidRev}, we demonstrate the existence of transient dynamics in the energy of linear perturbations of black hole spacetimes. Note that with the energy of a perturbation as the object of study, we are inescapably led to explore non-normality of the Hamiltonian with respect to an energy inner product, previously argued to be a natural norm for computing pseudospectra \cite{Jaramillo:2020tuu}. For previous discussions of the possibility of transients in the context of gravitational wave physics see \cite{Jaramillo:2022kuv, Boyanov:2022ark}.

\begin{figure}[h!]
\centering
\includegraphics[width=0.8\columnwidth]{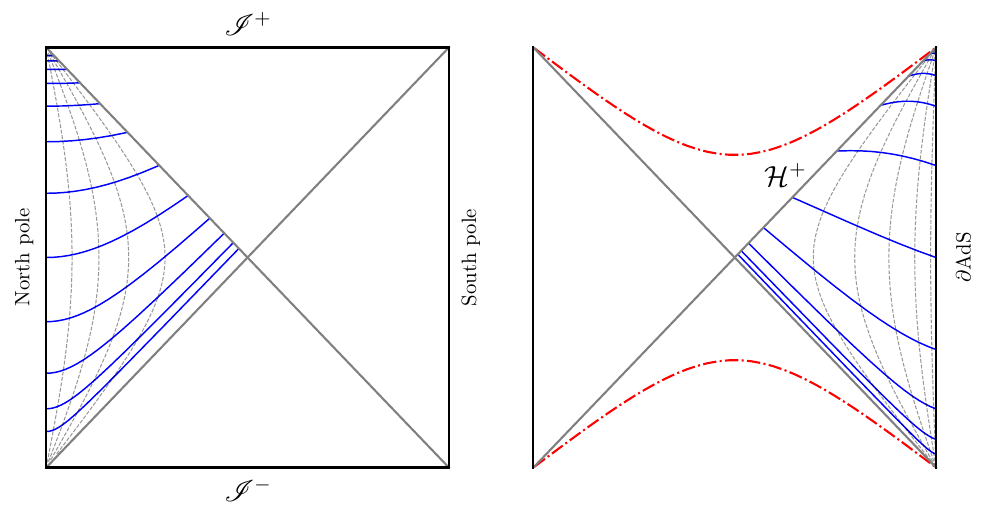}
\includegraphics[width=0.68\columnwidth]{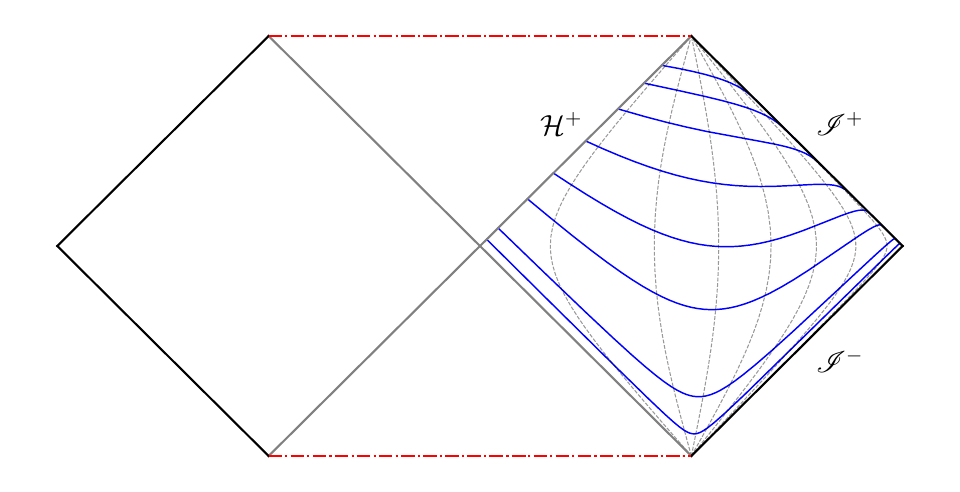}
\caption{Hyperboloidal slices used in this work (shown in blue), covering a de Sitter static patch (upper left), an exterior region of the Schwarzschild-AdS black brane (upper right) and an exterior region of Schwarzschild in asymptotically flat spacetime (lower). The Hamiltonian associated to evolution of linear perturbations on these slices is a non-normal operator with respect to the energy inner-product, and we study the associated transient effects through sums of its eigenfunctions, quasinormal modes.}
\label{fig:slices}
\end{figure}

We consider linear perturbations of cosmological spacetimes and black holes, studying transient behaviour by computing energies on hyperboloidal slices piercing their future horizons (see figure \ref{fig:slices}). Contrary to the Orr-Sommerfeld case, where the disturbance can draw energy from the mean flow even at a linear level, here energy can only leave the domain falling through the future horizon or escaping to future null infinity. Accordingly, the systems we study cannot exhibit transient \emph{growth}, but nevertheless we show that they do exhibit transient dynamics in the form of sums of QNMs that are arbitrarily long-lived.\footnote{In the AdS black brane case, however, one could consider an open holographic system with energy injected at infinity through a choice of boundary conditions. For instance, one could consider black brane spacetimes holographically dual to flow inside a channel. See for example \cite{Erdmenger:2018svl}. } In particular we show that there exist a sum of $M$ QNMs with lifetime $\sim \log{M}$. Such transients have a natural geometric interpretation as perturbations whose energy is localised near the future horizon and propagates along it (and also along future null infinity where applicable) -- indeed, non-normality is often associated to the existence of horizons in the first place, and so the existence of such special perturbations further solidifies this association. 

The structure of the paper is as follows. In section \ref{sec2_generalities} we introduce the general tools we employ in this paper. This includes the energy of perturbations on the leaves of a hyperboloidal foliation and the associated energy inner product. We also introduce techniques to construct an orthogonal basis for solutions built from a finite sum of QNMs, and how to use them to construct the longest-lived perturbations. In section \ref{dS_section} we apply these techniques to perturbations of the de Sitter static patch, where we proceed analytically. In section \ref{SAdS_section} we turn to scalar perturbations of the Schwarzschild-AdS black brane and its CFT dual, with an analytic example for BTZ and numerical results for higher dimensions. We examine the asymptotically flat Schwarzschild solution in section \ref{flat_section}, focusing on $l=2$ gravitational perturbations. We discuss the relation between transients and pseudospectra in section \ref{sec:pseudospectra}, where we also present the dS$_{d+1}$ static patch pseudospectra. We finish with a discussion in section \ref{sec:discussion}.

\section{Generalities}\label{sec2_generalities}
In this work we consider linear perturbations of fixed background spacetimes. We examine complex scalar perturbations governed by the equation of motion $\left(\Box - m^2\right)\Phi = 0$, describing the propagation of $\Phi$ on these fixed backgrounds. In the asymptotically flat case we also consider the standard treatment of scalar, electromagnetic and gravitational perturbations according to the Regge-Wheeler-Zerilli equations. In this section we give the definitions and techniques that we use throughout the rest of the paper to demonstrate the transient dynamics of linear perturbations for each spacetime.

\subsection{Hyperboloidal slices}
In general, perturbations can fall through the cosmological horizon of the static patch observer in the dS case, through the black hole event horizon in the Schwarzschild and Schwarzschild-AdS cases, and propagate to null infinity in Schwarzschild. In order to keep track of the amount of energy that leaves the region of interest (i.e. a static patch of dS and an exterior region of the black holes), it is convenient to work with a set of `hyperboloidal' slices \cite{Schmidt:1993rcx, Dyatlov:2010hq, Bizon:2010mp,  Warnick:2013hba, PanossoMacedo:2018hab, Gajic:2019qdd, Bizon:2020qnd, PanossoMacedo:2023qzp} -- spacelike slices that pierce the future horizons, as shown in figure \ref{fig:slices}. These are reached starting from Schwarzschild-like coordinates,
\be
\label{eq:lineelem_general}
ds^2 = -f(r)dt^2 + \frac{dr^2}{f(r)} + r^2 d\Sigma^{2}_{d-1},
\ee
via the following change of coordinates,
\be
\label{eq:hyperboloidal_general}
t = \tau - h(z), \qquad r = R(z).
\ee
Here the new radial coordinate $z \in [0,1]$ is chosen such that $z=1$ corresponds to the future horizon, while $z=0$ corresponds to the origin of coordinates at the North pole of dS, the conformal boundary of Schwarzschild-AdS, and future null infinity in the case of Schwarzschild. Full details of the functions $f(r), h(z), R(z)$ are given in later sections, specialised appropriately to each spacetime.

\subsection{Energy and QNMs}
For the complex scalar $\Phi$, the energy on the spacelike slice labelled by $\tau$, $\Sigma_\tau$, is given by
\be
\label{eq:total_energy_Phi}
E(\tau) =\int_{\Sigma_{\tau}} T^\mu_{\phantom{\mu}\tau}\,n_{\mu} d\Sigma_{\tau},
\ee
where $n = \left(-1/\sqrt{-g^{\tau\tau}}\right)\, d\tau$ is the unit, future-directed normal to $\Sigma_\tau$ and $T_{\mu\nu}$ is the stress-energy tensor of the scalar field,
\be
\label{eq:energymomentum_Phi}
T_{\mu\nu} = \frac{1}{2}\nabla_{\mu} \overline{\Phi}\nabla_{\nu}\Phi + \frac{1}{2}\nabla_{\nu}\overline{\Phi}\nabla_{\mu}\Phi - g_{\mu\nu}\(\frac{1}{2}\nabla_{\alpha}\overline{\Phi}\nabla^{\alpha}\Phi + \frac{1}{2}m^2\overline{\Phi}\Phi \),
\ee
where $\overline{\Phi}$ denotes the complex conjugate of $\Phi$.\footnote{An alternative way to derive \eqref{eq:total_energy_Phi} is starting from the Lagrangian density ${\cal L} = \sqrt{-g}(-\frac{1}{2}\nabla_\mu \overline{\Phi}\nabla^\mu \Phi -\frac{1}{2} m^2 \overline{\Phi}\Phi)$ in hyperboloidal coordinates. Then $E(\tau) = \int {\cal E} dz d\Sigma_{d-1}$ where ${\cal E}$ is the Hamiltonian density, ${\cal E} = \dot{\overline{\Phi}}\frac{\delta {\cal L}}{\delta \dot{\overline{\Phi}}} + \dot{\Phi}\frac{\delta {\cal L}}{\delta \dot{\Phi}} - {\cal L}$ where dots denote $\tau$ derivatives. The equation of motion derived from ${\cal L}$ is \eqref{eq:block_hamiltonian} where applying the differential operator $\mathcal{H}$ is equivalent to acting with $\mathcal{E}$ using the Poisson bracket (after accounting for field redefinitions). \label{footnote:generators}}
Note that $T^\mu_{\phantom{\mu}\tau}$ is the conserved current associated to the Killing vector $\partial_\tau$, and \eqref{eq:total_energy_Phi} tracks the associated amount of Noether charge on $\Sigma_\tau$. Without loss of generality, $\Phi$ can be decomposed into orthogonal eigenfunctions in the $d-1$ transverse directions, 
\bea
\Phi(\tau, z, \Omega_{d-1}) &=& \sum_{l \mathbf{m}}z^{\beta} \phi_l(\tau,z) Y_{l \mathbf{m}}(\Omega_{d-1}) , \label{eq:Phi_decomp_harmonics}\\
\Phi(\tau, z, \Vec{x}) &=& \int \frac{d^{d-1}k}{(2 \pi)^{d-1}}z^{\beta} \phi_{\Vec{k}}(\tau,z)e^{i \Vec{k}\cdot \Vec{x}} , \label{eq:Phi_decomp_planewaves}
\eea
where \eqref{eq:Phi_decomp_harmonics} is used for dS$_{d+1}$ and Schwarzschild, while \eqref{eq:Phi_decomp_planewaves} is used for planar Schwarzschild-AdS$_{d+1}$.
Here $\Omega_{d-1}$ denotes the angles on a $S^{d-1}$ sphere and $Y_{l \mathbf{m}}(\Omega_{d-1})$ are generalised spherical harmonics, with $l$ and $\mathbf{m}$ the total and azimuthal angular momentum quantum numbers. The plane wave momentum and transverse position vectors are denoted by $\Vec{k},\,\Vec{x} \in \mathbb{R}^{d-1}$, respectively. In addition, we have introduced a factor of $z^\beta$ which will be convenient later for enforcing the required behaviour at $z=0$. 

With decompositions of perturbations performed as above, the total energy \eqref{eq:total_energy_Phi} reduces to a sum of individual contributions from each $l$ and $\vec{k}$, due to orthogonality of the transverse eigenfunctions. For convenience, we thus consider the energy of a single $l$-mode, $\phi_l(\tau, z)$, or $\vec{k}$-mode, $\phi_{\vec{k}}(\tau, z)$, and drop the associated label: $\phi(\tau, z)$. The energy \eqref{eq:total_energy_Phi} for such an individual perturbation $\phi(\tau, z)$ is then given by
\be
E[\xi] = \left<\xi, \xi\right>, \label{eq:mode_energy}
\ee
where we have introduced the notation $\xi(\tau,z) \equiv \begin{pmatrix} \phi(\tau,z)\\ \partial_\tau \phi(\tau,z) \end{pmatrix}$, and the energy inner product
\be
\label{eq:energy_innerprod}
\left<\xi_1, \xi_2\right> = \frac{1}{2}\int_0^1 dz \(w(z) \,\p_{\tau}\overline{\phi}_1 \p_{\tau}\phi_2 + p(z) \,\p_z \overline{\phi}_1 \p_z \phi_2 + q(z) \,\overline{\phi}_1 \phi_2\).
\ee
The functions $w(z), p(z), q(z)$ are provided later for each case of interest.
See also \cite{Gasperin:2021kfv} for further discussions of energy inner products in hyperboloidal slicings. 
The flux of energy ${\cal F}$ leaving the region at time $\tau$ is given by taking the time derivative of \eqref{eq:mode_energy}, which in the case of dS$_{d+1}$ and Schwarzschild-AdS$_{d+1}$ cases reduce to a boundary term at the future horizon ($z=1$),
\be
{\cal F}(\tau) \equiv -\partial_\tau E[\xi(\tau,z)]= |\p_{\tau}\phi|^2 \Big|_{z=1},\label{eq:FluxCC}
\ee
while in the asymptotically flat case there is also a flux at future null infinity ($z=0$),
\be
{\cal F}(\tau) \equiv -\partial_\tau E[\xi(\tau,z)]= |\p_{\tau}\phi|^2 \Big|_{z=1} + |\p_{\tau}\phi|^2 \Big|_{z=0}.\label{eq:FluxFlat}
\ee
Thus, ${\cal F}(\tau) \geq 0$ and $E(\tau)$ is a non-increasing function of $\tau$; energy cannot enter the region, only leave through the future horizon, propagate to future null infinity, or stay the same.\footnote{Note that if we choose $m^2 < 0$ in dS$_{d+1}$ there are unstable, exponentially growing perturbations. In such a case we still have $\partial_\tau E(\tau) \leq 0$, but now $E(\tau)$ is unbounded from below ($q(z)$ becomes negative in \eqref{eq:energy_innerprod}). We do not consider unstable systems in this work.}

We now arrive at the crucial physical feature: the Hamiltonian ${\cal H}$ describing the $\tau$ evolution of the scalar perturbation is a non-normal operator with respect to the energy inner product \eqref{eq:energy_innerprod}. Here ${\cal H}$ is obtained by a first-order reduction of the Klein-Gordon equation,
\be
\label{eq:block_hamiltonian}
i \partial_\tau \xi = {\cal H} \xi , \qquad \mathcal{H} = \begin{pmatrix} 0 & i \\ \mathcal{L}_1 & \mathcal{L}_2 \end{pmatrix},
\ee
where the differential operators $\mathcal{L}_1$, $\mathcal{L}_2$ are given in terms of the functions appearing in \eqref{eq:energy_innerprod}, 
\bea
\mathcal{L}_1 &=& \frac{i}{w(z)}\left( p(z) \p^2_z + p'(z)\p_z - q(z) \right), \label{eq:diffop_1_general}\\
\mathcal{L}_2 &=& \frac{i}{w(z)}\left( 2 \gamma(z) \p_z + \gamma'(z)\right), \label{eq:diffop_2_general}
\eea
defining $\gamma(z) = h'(z) \,p(z)$. Note that time translations, as generated by $\mathcal{H}$, are equivalently generated by the energy (see footnote \ref{footnote:generators}). We demonstrate non-normality in appendix \ref{sec:non_normality}. In particular, the (regular, normalisable) eigenfunctions of ${\cal H}$ are not orthogonal with respect to the energy inner product \eqref{eq:energy_innerprod}.

Such eigenfunctions, with their exponential time dependence fixed by equation \eqref{eq:block_hamiltonian} $\xi_n(\tau,z) = e^{-i \omega_n \tau} \Tilde{\xi}_n(z)$, labelled by $n$, correspond to scalar QNMs of the spacetime \eqref{eq:lineelem_general}; the eigenvalues $\omega_n$ being their associated QNM frequencies. Consequently, a perturbation formed from a sum of QNMs,
\be
\label{eq:qnm_sum}
\xi(\tau, z) = \sum_{n=1}^{M}c_n e^{-i \omega_n \tau} \Tilde{\xi}_n(z),
\ee
has an energy \eqref{eq:mode_energy} which is not simply a sum of the energies of each QNM,
\bea
E[\xi] &=& \sum_{n=1}^{M}\sum_{m=1}^{M} c_n^*c_m e^{i (\omega_n^*-\omega_m) \tau} \left<\Tilde{\xi}_n, \Tilde{\xi}_m\right>\\
&=& \sum_{n=1}^{M}|c_n|^2 e^{2\im{\,\omega_n}\tau} E[\Tilde{\xi}_n] + \text{cross-terms},
\eea
because of the off-diagonal/cross-terms, which arise due to non-orthogonality of QNMs under the energy inner-product. The appearance of these cross terms opens the door to transient dynamics, since without them, the slowest possible decay of energy is simply determined by the longest-lived QNM, i.e. obtained here by setting $c_n = 0$ for all $n$ except for the QNM(s) with the largest $\im{\,\omega_n}$.

\subsection{Energy growth curve}
\label{sec:energy_growth}
In the preceding sections we established that energy outside horizons does not have to decay according to QNM decay rates, because of the non-normality of ${\cal H}$ with respect to the inner product \eqref{eq:energy_innerprod}. We also established that energy cannot grow. This leads to the question, given a sum of QNMs, what is the slowest that energy can decay? The techniques we use here parallel those given in \cite{Reddy93} for the Orr-Sommerfeld case.\footnote{For a similar application of the techniques to a discrete Hamiltonian with localised losses see \cite{PhysRevA.98.052105}. In contrast with this, and the Orr-Sommerfeld case, our main results rely on dynamics governed by event horizons.}

We can compute the slowest possible decay using the so-called \emph{energy growth}\footnote{This name is inherited from the hydrodynamics literature where, as we outlined for the Orr-Sommerfeld case, energy can grow.} curve, which determines the maximum possible energy at a specific moment in time $\tau$ relative to the energy at a fiducial initial time $\tau = 0$, defined as
\bea
G(\tau) &\equiv& \sup_{\xi(0,z)} \frac{E[\xi(\tau,z)]}{E[\xi(0,z)]}\\ 
&=& \sup_{\xi(0,z)} \frac{E[e^{-i \mathcal{H}\tau}\xi(0,z)]}{E[\xi(0,z)]} \label{eq:maxresponse_2}.
\eea
In the second line, we have used that $\xi(\tau,z) = e^{-i \mathcal{H}\tau} \xi(0,z)$ is a formal solution to \eqref{eq:block_hamiltonian}. Note that \eqref{eq:maxresponse_2} is, by definition, the norm-squared of the time evolution operator, thus we may write
\be
G(\tau) = \|e^{-i \mathcal{H}\tau}\|^{2}_{E} \,.
\ee
Also note that $G(\tau)$ is upper-bounded by $1$ since $G(0) = 1$ and $\partial_\tau E \leq 0$, while a simple lower bound on $G(\tau)$ follows immediately from the existence of the longest-lived QNM. In other words, $G(\tau)$ obeys the two-sided bounds,
\be
e^{2\im{\,\omega_0}\tau} \leq G(\tau) \leq 1, \qquad \forall \tau \geq 0,\label{Gbounds}
\ee
where $\omega_0$ denotes the frequency of the longest-lived QNM.

Let us denote the subspace of solutions generated by a sum of $M$ QNMs (as in \eqref{eq:qnm_sum}) as $W$. Restricting to this subspace we can then find the energy growth by projecting the Hamiltonian $\mathcal{H}$ onto $W$, $\mathcal{H}_W$, such that it satisfies $\mathcal{H}\zeta=\mathcal{H}_W \zeta$ for any $\zeta \in W$. The energy growth in $W$ is then
\be
\label{eq:energygrowth_subspace_W}
G_W(\tau)\equiv \sup_{\xi(0,z) \in W} \frac{E[\xi(\tau,z)]}{E[\xi(0,z)]} = \|e^{-i \mathcal{H}_W\tau}\|^{2}_{E} \,.
\ee
Note that since $W$ is a subspace of solutions, $G_W(\tau)$ provides a lower bound on $G(\tau)$
\be
e^{2\im{\,\omega_0}\tau} \leq G_W(\tau) \leq G(\tau) \leq 1, \label{GWboundsG}
\ee
where the leftmost inequality holds provided $W$ contains the longest lived QNM. Also note that in the dS$_{d+1}$ and Schwarzschild-AdS$_{d+1}$ cases, $G_W(\tau)$ depends only on the spectrum of the theory. This can be seen if one normalises QNMs at the horizon, without loss of generality, such that $\phi_n(\tau,1) = e^{-i\omega_n\tau}$. Then \eqref{eq:FluxCC} can be computed for a sum of QNMs with arbitrary coefficients $c_n$, integrated to obtain $E(\tau)$, and $G_W(\tau)$ obtained by exploring the space of $c_n$.

In order to compute \eqref{eq:energygrowth_subspace_W} in this work, we first find an orthonormal basis for the subspace $W$. This can be done applying the Gram-Schmidt process with respect to the energy inner product to the set of $M$ QNMs. We first normalise the QNM spatial eigenfunctions $\{\Tilde{\xi}_n(z)\}_{n=1}^{M}$ to one, and then the Gram-Schmidt process produces the orthonormal set $\{\psi_n(z)\}_{n=1}^{M}$ satisfying $\langle \psi_i, \psi_j\rangle = \delta_{ij}$, so that initial data in $W$ can be expressed as follows,
\be
\label{eq:xiexpansion_M_orthbasis}
\xi(0, z) = \sum_{n=1}^{M}c_n\tilde{\xi}_n(z) = \sum_{n=1}^{M} d_n \psi_n(z) .
\ee
Given some initial data $\xi(0, z)$, it is now possible to extract the coefficients $d_n$ in the expansion above in a straightforward way as $d_n=\langle \psi_n, \xi(0, z) \rangle$, and then relate back to the coefficients $c_n$ in the eigenfunction basis via the $M\times M$ transformation matrix $U_W$ coming from the Gram-Schmidt process, $\Vec{d}=U_W \Vec{c}$. 

Armed with this new basis of orthonormal functions spanning $W$, one can prove that (see appendix \ref{sec:HWproof}) 
\be
\|\mathcal{H}_W\|_{E}^{2}=\|H_W\|_{2}^{2} \qquad \text{where} \qquad H_W \equiv U_W D_W U_{W}^{-1} , \label{HWdef}
\ee
with $D_W = \text{diag}(\omega_1,\omega_2,\ldots,\omega_M)$, and where $\|\cdot\|_2$ denotes the usual $l^2$-norm induced from the Euclidean inner product $\langle \Vec{e}_1, \Vec{e}_2 \rangle_2 = \Vec{e}_{1}^{\,\,*}\Vec{e}_{2}$ -- here and throughout this paper, $^*$ denotes the conjugate transpose. We can therefore express the maximum response in energy at a time $\tau$ to optimal initial data restricted to the finite subspace of QNMs $W$ as
\be
\label{eq:GW_2norm}
G_W(\tau)=\|e^{-i H_W\tau}\|^{2}_{2} \,.
\ee
Computing $G_W(\tau)$ is now straightforward since the $l^2$-norm of a matrix is given by its maximum singular value. Furthermore, not only can we determine the maximum energy at a given time $\tau$, but we can also extract the coefficients $\Vec{d}$ in $\eqref{eq:xiexpansion_M_orthbasis}$ for the initial data that results in this maximum response at $\tau$. They correspond to the entries in the right principal singular vector in the singular value decomposition of $e^{-i H_W\tau}$. We refer to a solution obtained in this way as an `optimal perturbation' in what follows.

Finally, it is important to reiterate that the energy growth $G_W(\tau)$ is not the time evolution of the energy of a perturbation. Rather, there exists initial data for each $\tau \geq 0$, $\xi(0, z) \in W$ with $E[\xi(0, z)]=1$, such that $E[\xi(\tau,z)]=G_W(\tau)$.

\section{de Sitter static patch}\label{dS_section}

In this section we focus on the static patch of de Sitter corresponding to the causal diamond for an observer sitting at the North pole, with associated future and past cosmological horizons \cite{PhysRevD.15.2738}. It is then possible to study linear perturbations of the static patch in the same fashion as one does for the exterior of a black hole, by constructing perturbations that are regular on the future cosmological horizon \cite{Lopez-Ortega:2006aal, Gibbons:2007fd}. Because dS$_{d+1}$ is a maximally symmetric spacetime whose isometry algebra $\mathfrak{so}(d+1,1)$ is the conformal algebra in $\mathbb{R}^d$, its QNMs are organised into conformal families and are analytically tractable.

In Schwarzschild-like coordinates the static patch takes the form \eqref{eq:lineelem_general} where the transverse geometry is the unit round $d-1$ sphere, $d\Sigma_{d-1}^2 \equiv d\Omega_{d-1}^2$ and the metric function
\be
f(r)=1-r^2,
\ee
where we have set the Hubble constant $H = L_{dS}^{-1} = 1$, and all times and frequencies in this section will be given in Hubble units. The hyperboloidal slices shown in figure \ref{fig:slices}, on which we are to compute the energy of the scalar field, are given by constant $\tau$ hypersurfaces in the coordinates \eqref{eq:hyperboloidal_general} with
\bea
h(z) &=& \frac{1}{2}\log (1-z), \\
R(z) &=& \sqrt{z}\,.
\eea
These slices are a generalisation of \cite{Bizon:2020qnd} to higher $d$. In these coordinates, at fixed $\tau$, $z=1$ corresponds to the future cosmological horizon while $z=0$ corresponds to the North pole. For the field redefinition \eqref{eq:Phi_decomp_harmonics} we make the parameter choice $\beta = l/2$. The functions defining the energy inner product \eqref{eq:energy_innerprod} and the differential operators in the Hamiltonian \eqref{eq:diffop_1_general} and \eqref{eq:diffop_2_general} read
\bea
w(z) &=& \frac{1}{2}\frac{z^{l+d/2}}{z}, \label{eq:dS_w}\\
p(z) &=& 2\,(1-z)\,z^{l+d/2}, \\
q(z) &=& \frac{1}{2}\(l(d+l) + m^2\)\frac{z^{l+d/2}}{z}, \label{eq:dS_q}\\
\gamma(z) &=& -z^{l+d/2}.
\eea

The static patch QNMs are $\xi^\pm_{nl} = (\phi^\pm_{nl}, \partial_\tau \phi^\pm_{nl})^T$ with
\bea
\phi^\pm_{nl}(\tau, z) &=& e^{-i \omega^\pm_{nl} \tau} {}_2F_1\left(-n,\frac{d}{2} - n - \Delta_\pm; \frac{d}{2}+l; z\right),\label{eq:dS_QNMs}\\
\omega^\pm_{nl} &=& -i \(\Delta_\pm + 2n + l\),\label{eq:dS_QNFs}
\eea
where $\Delta_\pm$ are the two roots of $-\Delta(\Delta - d) = m^2$. Note that both roots are valid QNMs, in contrast to the asymptotically AdS case considered later. When $m^2 \leq d^2/4$ the QNM frequencies are purely imaginary, i.e. $\omega^\pm_{nl} = -(\omega^\pm_{nl})^*$, and when $m^2 > d^2/4$ the modes appear in pairs related by reflections in the imaginary axis, $\omega^\pm_{nl} = -(\omega^\mp_{nl})^*$. The dilatation generator of the Euclidean conformal algebra, $D$, corresponds to the Hamiltonian, ${\cal H}$. In other words, for each root, \eqref{eq:dS_QNMs} correspond to a conformal family, with \eqref{eq:dS_QNFs} their conformal dimensions. These QNMs correspond to sources on the past conformal boundary at the North pole and on the future conformal boundary at the South pole, and a singular field configuration propagating between them along the past horizon of the static patch.\footnote{This is most easily seen by constructing e.g. the conformal primary wavefunction in global coordinates, starting from embedding space.}

We are now ready to compute energy growth curves for $W$, the subspace of linear perturbations comprised of $M$ QNMs. Prior to this, we provide a simple analytic (but non-optimal) example of a solution in $W$ whose lifetime grows monotonically with $M$.

\subsection{An analytic example}
As a simple analytic demonstration of the effects of non-normality, consider the following scalar field perturbation
\be
\phi(\tau, z) = a_1 \phi_1(\tau, z) + a_2 \phi_2(\tau, z),
\ee
formed from two $l=0$ QNMs of the static patch of dS$_4$,
\be
\phi_1(\tau, z) = \phi^-_{00}(\tau, z) =e^{-\tau}, \qquad \phi_2(\tau, z) = \phi^+_{00}(\tau, z)= e^{-2\tau}. \label{dS_twomodes}
\ee
These correspond to $\Delta = 1, n=0$ and $\Delta = 2, n=0$ perturbations at $m^2 = 2$ respectively. The energy of this perturbation \eqref{eq:total_energy_Phi} is
\be
E(\tau) = \frac{1}{2}|a_1|^2 e^{-2\tau} + \frac{2}{3}(a_1^*a_2 + a_1a_2^*)e^{-3\tau} + |a_2|^2 e^{-4\tau}.\label{eq:energy2modes}
\ee
Crucially, note the appearance of the cross-term. The cross-term is a direct consequence of the non-orthogonality of the modes \eqref{dS_twomodes} with respect to the inner product \eqref{eq:energy_innerprod}. Without this term, the perturbation exhibiting the slowest decay of energy would be obtained simply by setting $a_2 = 0$, and the associated lifetime given by the longest-lived QNM, i.e. $E \sim (e^{-\tau})^2$. However, in this non-normal system, the existence of the cross-term means that the QNM lifetime does not set the slowest possible decay. In particular, in the neighbourhood of $\tau = 0$ one has
\be
\frac{E(\tau)}{E(0)} = 1 - 6\frac{|a_1+2a_2|^2}{3|a_1|^2 + 6 |a_2|^2 + 4(a_1^*a_2 + a_1 a_2^*)}\tau + O(\tau)^2.
\ee
Note that the coefficient of $\tau$ appearing in this Taylor series is necessarily non-positive because the energy cannot grow. Choosing $a_2 = -a_1/2$ removes the $O(\tau)$ (and also $O(\tau)^2$) contributions and thus delays the decay of the energy.

This process can be continued indefinitely; by adding more modes in the QNM sum, one can further delay the onset of modal energy decay as follows. We begin by enumerating all of the $d=3$, $l=0$, $m^2 = 2$ QNMs together by a single integer $k = 1,2, \ldots$. For $k$ odd we pick a $\Delta = 1$ QNM with $n=(k-1)/2$, while for $k$ even we pick a $\Delta = 2$ QNM with $n = (k-2)/2$, 
\be
\phi_k(\tau, z) = e^{-k \tau}\frac{(1 + \sqrt{z})^k - (1-\sqrt{z})^k}{2k \sqrt{z}}, \qquad k = 1,2,\ldots\,.
\ee
We then construct a perturbation from a sum the first $M$ modes as follows,
\bea
\phi(\tau, z) &=& \sum_{k = 1}^M (-1)^{1+k} 2^{\frac{3}{2}-k} \sqrt{M(2M-1)}\frac{\Gamma\left(M\right)}{\Gamma(k) \Gamma\left(1-k+M\right)} \phi_{k}(\tau, z)\\
&=& 2^{\frac{1}{2}-M} \sqrt{\frac{2M-1}{M}} \frac{\left(2 + e^{-\tau}(\sqrt{z} - 1)\right)^M-\left(2 - e^{-\tau}(\sqrt{z} + 1)\right)^M}{\sqrt{z}}.\label{examplephisol}
\eea
The coefficients in this sum have been chosen to remove the first $2M-2$ powers of $\tau$ in the Taylor expansion of $E(\tau)$ for this perturbation.\footnote{For a related construction in a different context, see \cite{PhysRevA.98.052105}.} The energy is given by
\bea
E(\tau) &=& 1 - (1-e^{-\tau})^{2M-1} (1 + (2M-1)e^{-\tau})\\
&=& 1 - 2M\tau^{2M-1} + O(\tau)^{2M},
\eea
with an outgoing energy flux \eqref{eq:FluxCC} given by ${\cal F}(\tau) = 2M(2M-1) e^{-2\tau} (1-e^{-\tau})^{2M-2}$ peaked with a maximum at $\tau = \log M$. Further examination of the perturbation reveals that at $\tau = 0$ energy is localised near the future horizon, propagating outwards from the horizon in subsequent evolution. Increasing $M$ leads to a larger maximum of the initial energy density, resulting in a longer decay time. Note that the solution \eqref{examplephisol} is real, and thus is also a solution in the case of a real scalar field.

This example proves, at least for $d=3$, $l=0$, $m^2 = 2$, that for any $\tau \geq 0$ one can construct $W$ containing perturbations such that $E(\tau)$ is arbitrarily close to one (by using a suitably large $M$). Through \eqref{GWboundsG} this implies $G(\tau) = 1$ in this case.
In the limit $M\to \infty$, \eqref{examplephisol} corresponds to a perturbation whose energy remains inside the static patch, since
\be
\lim_{M \to \infty} E(\tau) = 1,\qquad \lim_{M \to \infty} {\cal F}(\tau)  = 0\qquad (\tau \geq 0).
\ee
However, in this limit the spatial gradients of $\phi$ diverge at the horizon. Thus, we should consider only finite $M$ and, consequently, \eqref{examplephisol} is not inconsistent with an ultimate decay via individual QNM rates at asymptotically late times.

\subsection{Energy growth and optimal perturbations} \label{sec:dS_energygrowth}
The analytic example in the previous section demonstrates that the energy of a sum of QNMs can remain in the static patch for an arbitrarily long time (the lifetime grows logarithmically with $M$, the number of modes). Going further, in this section we design coefficients of a sum of $M$ QNMs to maximise the energy at any desired value of $\tau$ -- an optimal perturbation -- following the recipe outlined in section \ref{sec:energy_growth}.

Orthogonalising the finite set of $M$ QNMs with respect to the energy inner product \eqref{eq:energy_innerprod} constitutes the first step. For each chosen QNM at $\tau = 0$, given by \eqref{eq:dS_QNMs}, we construct the normalised eigenfunction, 
\be
\tilde{\xi}^\pm_{nl} = \frac{\xi_{nl}^\pm(0,z)}{\sqrt{\left<\xi_{nl}^\pm(0,z),\xi_{nl}^\pm(0,z)\right>}}. \label{dSGSnormalised}
\ee
For simplicity of notation, let us label the finite set of modes among \eqref{dSGSnormalised} with a single index, $\{\tilde{\xi}_n\}$, $n = 1, \ldots, M$. An orthonormal set of functions $\{\psi_n\}$ is then constructed following the Gram-Schmidt process, starting with an arbitrary identification $\psi_1 = \tilde{\xi}_1$. For illustration, the next function $\psi_2$ is constructed as
\be
\psi_2 = \frac{f_2}{\sqrt{\left<f_2,f_2\right>}},\qquad \text{where}\quad f_2 \equiv \tilde{\xi}_2 - \left<\psi_1, \tilde{\xi}_2\right> \psi_1
\ee
so that $\left<\psi_2, \psi_2\right> = 1$ and $\left<\psi_1, \psi_2\right> = 0$. The process continues in this fashion following Gram-Schmidt. The $U_W$ matrix that relates the expansion coefficients in \eqref{eq:xiexpansion_M_orthbasis} is then given by the appropriate projections between these two sets of functions, i.e. 
\be
U_W = \begin{pmatrix}
    \left<\psi_1, \tilde{\xi}_1\right> & \left<\psi_1, \tilde{\xi}_2\right> & \ldots & \left<\psi_1, \tilde{\xi}_M\right>\\
    0 & \left<\psi_2, \tilde{\xi}_2\right> & \ldots & \left<\psi_2, \tilde{\xi}_M\right>\\
    \vdots & \vdots & \ddots & \vdots\\
    0 & 0 & \ldots & \left<\psi_M, \tilde{\xi}_M\right>
\end{pmatrix}.
\ee
With $U_W$ constructed, the matrix $H_W$ is easily obtained according to \eqref{HWdef}. The results of interest then follow from the singular value decomposition of the $W$-projected time evolution operator, $e^{-i H_W \tau}$. Its maximum singular value computes the energy growth curve $G_W(\tau)$ following \eqref{eq:GW_2norm}, while its right principal singular vector gives the $\vec{d}$ vector for optimal initial data maximising the energy at time $\tau$ (i.e. initial data for a perturbation whose energy equals $G_W(\tau)$ at time $\tau$).

Note that all of the steps outlined just above are analytic in this case. This is because the QNMs are given in closed form \eqref{eq:dS_QNMs}, where the hypergeometric series truncates at finite order. Thus, the inner product between any two eigenfunctions can be computed by the finite sum,
\bea
\left< \xi^{\pm_1}_{n_1l}, \xi^{\pm_2}_{n_2l}\right> = \sum_{j_1 = 0}^{n_1}\sum_{j_2 = 0}^{n_2} \Bigg[ && \frac{(-n_1)_{j_1}\left(\frac{d}{2} - n_1 -\Delta_{\pm_1}^*\right)_{j_1}}{j_1!\left(\frac{d}{2}+l\right)_{j_1}} \frac{(-n_2)_{j_2}\left(\frac{d}{2}- n_2 - \Delta_{\pm_2}\right)_{j_2}}{j_2!\left(\frac{d}{2}+l\right)_{j_2}}\times\\
&&  \frac{m^2 + l(l+d) + (l+2n_1 + \Delta_{\pm_1}^*)(l+2n_2+\Delta_{\pm_2}) +\frac{8 j_1 j_2}{d+2l+2j_1+2j_2-2}}{2(d+2l + 2j_1+2j_2)}\Bigg],\nonumber
\eea
where $(x)_n$ are Pochhammer symbols and the left hand side is at $\tau = 0$.
In practice, the exact expressions are too slow to work with, and instead we utilise a floating point representation with 8000 digits of precision. The high degree of precision is required for a couple of reasons. Firstly, the Gram-Schmidt process itself is sensitive to round-off errors. Secondly, the optimal solutions we construct from the singular value decomposition of $e^{-i H_W \tau}$ rely on large cancellations in the initial data. 

\begin{figure}[h!]
\centering
\includegraphics[width=\columnwidth]{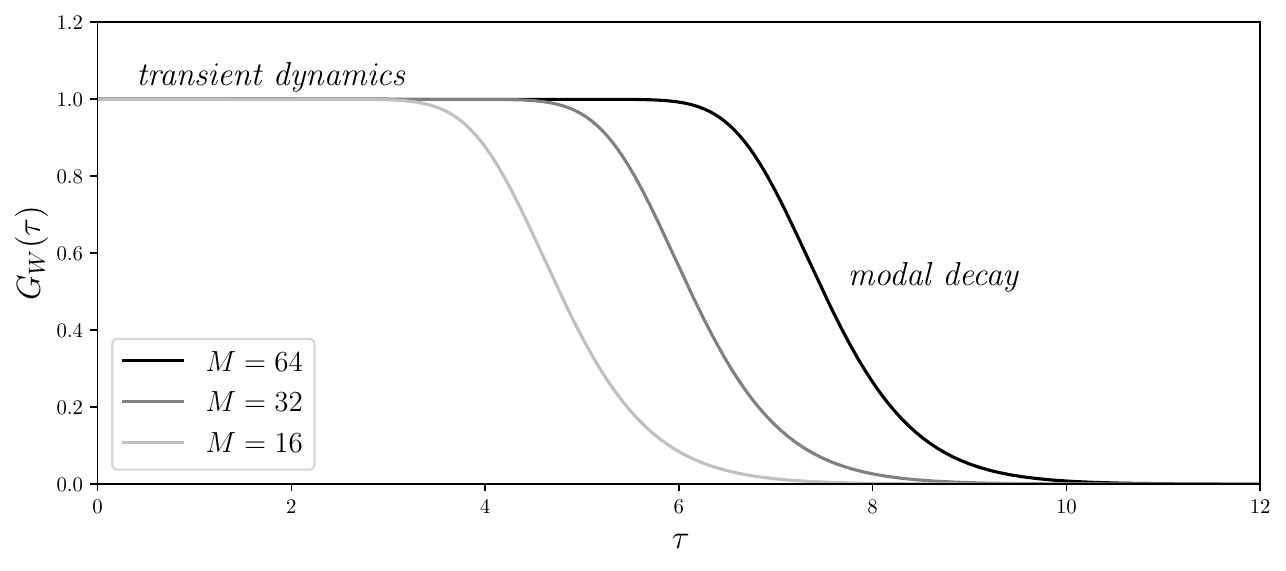}
\caption{
The energy growth $G_W(\tau)$ for scalar field perturbations of the static patch of de Sitter built from a sum of $M$ QNMs. For each $M$ shown, $G_W(\tau)$ is the maximum possible energy of the perturbation at time $\tau$, given unit initial energy, for the spatial slices shown in figure \ref{fig:slices}. For any given $\tau$, initial data can be constructed that has an energy $G_W(\tau)$ at time $\tau$. Because QNMs are eigenfunctions of a non-normal operator, $G_W(\tau)$ does not simply decay with the longest-lived mode, but rather exhibits transient dynamics as indicated. $G_W(\tau)$ shows the longest-lived transient behaviour at a given $M$. Increasing $M$ corresponds to longer and longer lived transients corresponding to linear perturbations that are localised near the cosmological horizon. Parameters used: $m^2 = 2$, $d=3$, $l=0$. Units are in Hubble times.}
\label{fig:dS_trend}
\end{figure}

The results are as follows. Figure \ref{fig:dS_trend} shows the $G_W(\tau)$ for the parameter choice $m^2=2$, $d=3$, $l=0$, for the three choices of the dimension of $W$, $M=16, 32, 64$. With these parameters the eigenfrequencies $\omega^\pm_{nl}$ are imaginary, and our labelling picks the first $M$ modes in order of decreasing $\im\,\omega^\pm_{nl}$. The most striking feature is the initial flat plateau of $G_W(\tau)$ -- a direct consequence of the non-normality of $\mathcal{H}$. Recall that $G_W(\tau)$ gives the maximum possible energy that can be formed in $W$ at a time $\tau$. Thus, a flat $G_W(\tau)$ curve means that one can construct a perturbation whose energy does not decay for the same amount of time. The maximum duration of a transient increases as $M$ is increased, with a growth in duration proceeding as $\sim \log M$. Note that logarithmic growth of the transient regime is consistent with the behaviour seen in the closed-form solution of the previous section, \eqref{examplephisol}. Also consistent with \eqref{examplephisol} is the behaviour $1-G_W(\tau) \sim \tau^{2M-1}$ for small $\tau$. Finally, the transient period ends and the decay of $G_W(\tau)$ reflects the slowest possible single-QNM decay, i.e. the decay rate of the fundamental mode.

\begin{figure}[h!]
\centering
\includegraphics[width=\columnwidth]{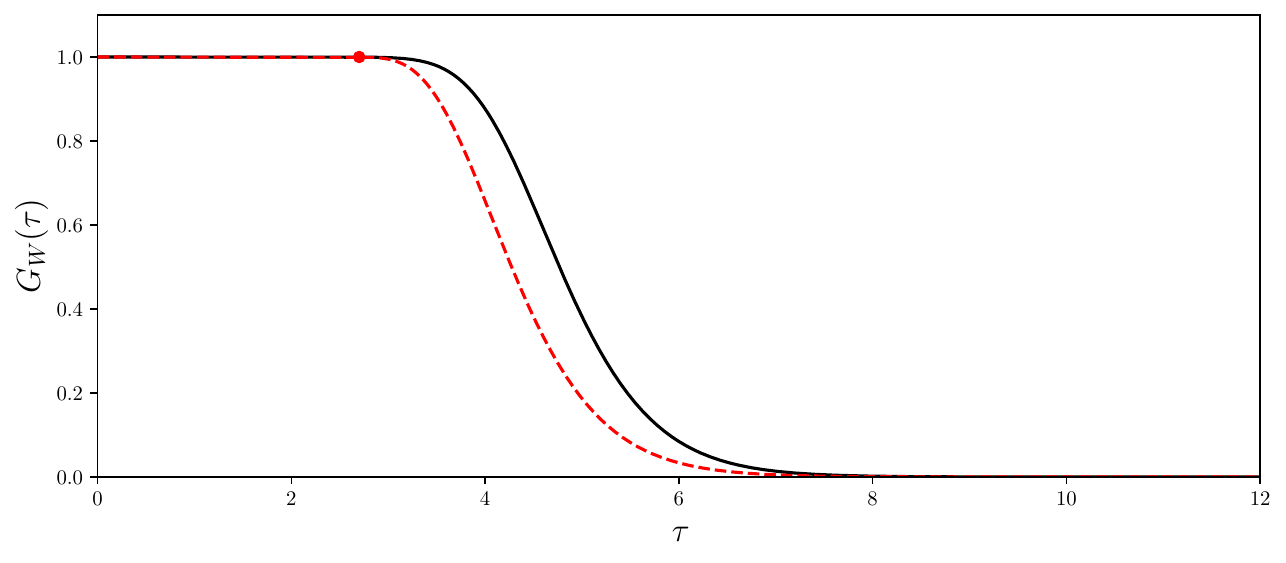}
\includegraphics[width=0.8\columnwidth]{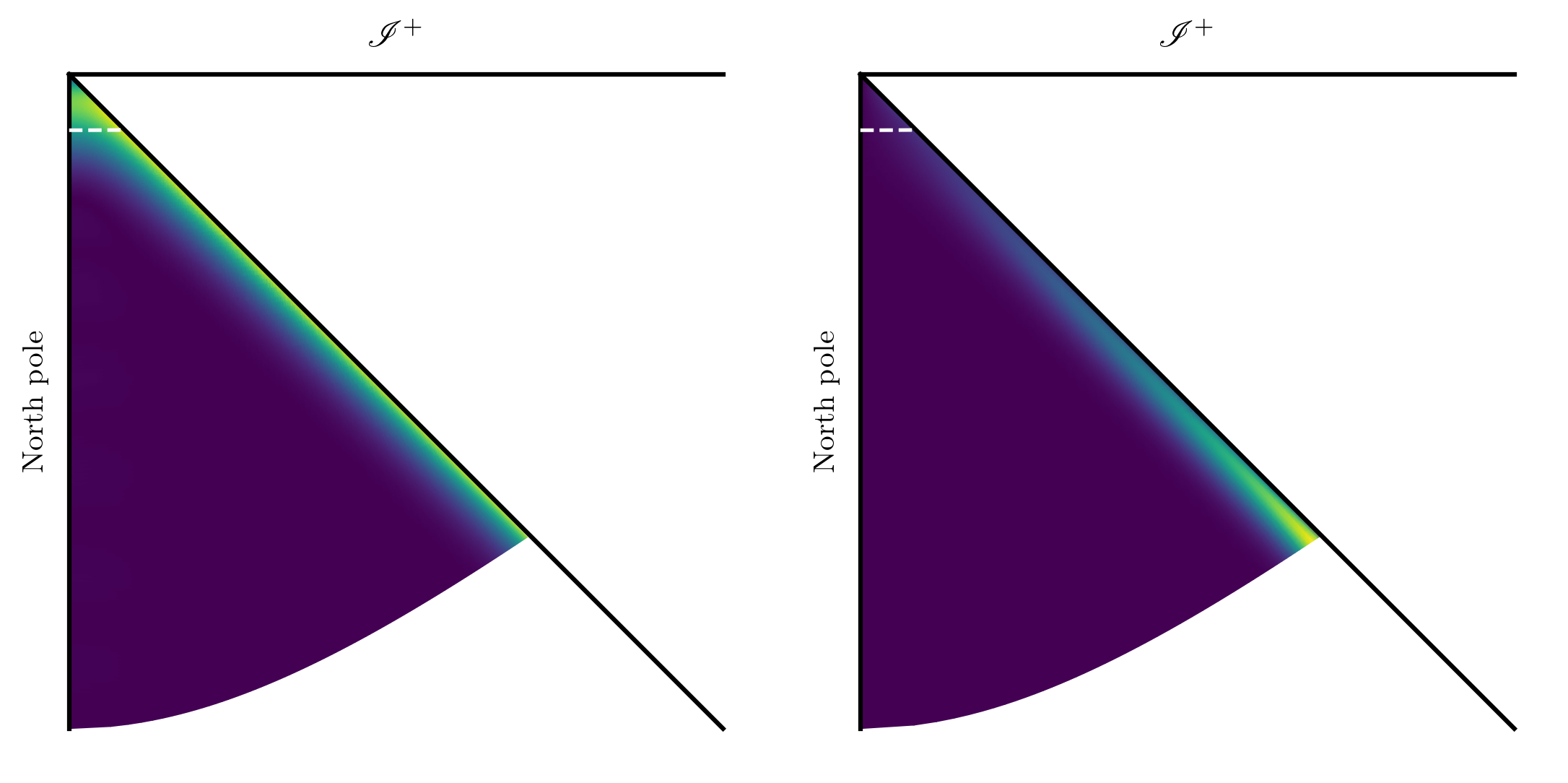}
\caption{An optimal sum of 16 QNMs in the static patch of de Sitter. Here the QNMs maximise the energy at a chosen time $\tau_* = 2.7$, computed from the singular value decomposition of the time evolution operator projected to the subspace of $16$ QNMs. Parameters used: $m^2 = 2$, $d=3$, $l=0$. \textbf{Upper panel:} The energy growth curve $G_W(\tau)$ (black), with the energy of the optimal perturbation for $\tau_* = 2.7$ (red dash). The red dot marks the time $\tau_* = 2.7$. Units are in Hubble times. \textbf{Lower left panel:} Colour indicates the value of $|\phi|$ for the optimal perturbation, shown in the dS conformal diagram. \textbf{Lower right panel:} Colour indicates the energy density of the optimal perturbation in the dS conformal diagram. The energy is localised near the cosmological horizon and gets closer to the horizon for larger values of $M$. The dashed white line is the spatial slice at $\tau_* = 2.7$, approximately where the energy begins its modal decay.}
\label{fig:dS_pick}
\end{figure}

To illustrate the optimal perturbation which tracks the transient plateau of $G_W(\tau)$ we pick a target time $\tau=\tau_*$ just before the plateau ends. An example is shown in figure \ref{fig:dS_pick} at the red dot, $\tau_* = 2.7$ for the growth curve at $M=16$. Using the singular value decomposition of $e^{-i H_W \tau_*}$, we then extract the coefficients $\vec{d}$ for the initial data in the orthonormal basis whose energy will satisfy $E[\xi(\tau_*,z)]=G_W(\tau_*)$. This is then converted back to coefficients in a sum of QNMs via $\vec{c} = U_W^{-1}\vec{d}$, and evolve this sum according to their eigenfrequencies. The resulting energy of the evolution is superposed on top of the growth curve in figure \ref{fig:dS_pick}. As anticipated, in order to reach $G_W(\tau)$ at $\tau=\tau_*$, its energy must remain constant until that time. The optimal perturbation is thus characterised by a long lifetime as compared with the lifetime of the slowest decaying QNM, and can be made arbitrarily long by increasing $M$.

In the lower panels of figure \ref{fig:dS_pick}, we examine the geometric nature of the $\tau_* = 2.7$ optimal perturbation in the conformal diagram of the dS$_4$ static patch. On the left, $|\phi|$ is displayed, showing how the optimal perturbation is peaked near the cosmological horizon, propagates along it, until it starts falling through at $\tau \simeq \tau_*$, corresponding to the white dashed time slice. In addition, on the right we plot the energy density of this solution as given by the $z$-integrand in \eqref{eq:energy_innerprod}, observing an energy packet travelling just outside and along the horizon. The energy density decays as the packet disperses, but the total energy on the slice remains approximately constant until $\tau_*$. This highlights the link between causal horizons and non-normality; non-normality is necessary for a long-lived QNM sum, and the geometrical reason it can remain long lived is because of its ability to propagate along the horizon for an arbitrarily long time.

Perturbations with nonzero angular momentum quantum numbers $l,\mathbf{m}$ display similar behaviour, but the transient plateau is shorter. Since different $l,\mathbf{m}$ modes are orthogonal, the $l = \mathbf{m} = 0$ modes we have constructed here therefore provide the optimal contributions, even when allowing for nontrivial profiles in the transverse directions.

\begin{figure}[h!]
\centering
\includegraphics[width=0.8\columnwidth]{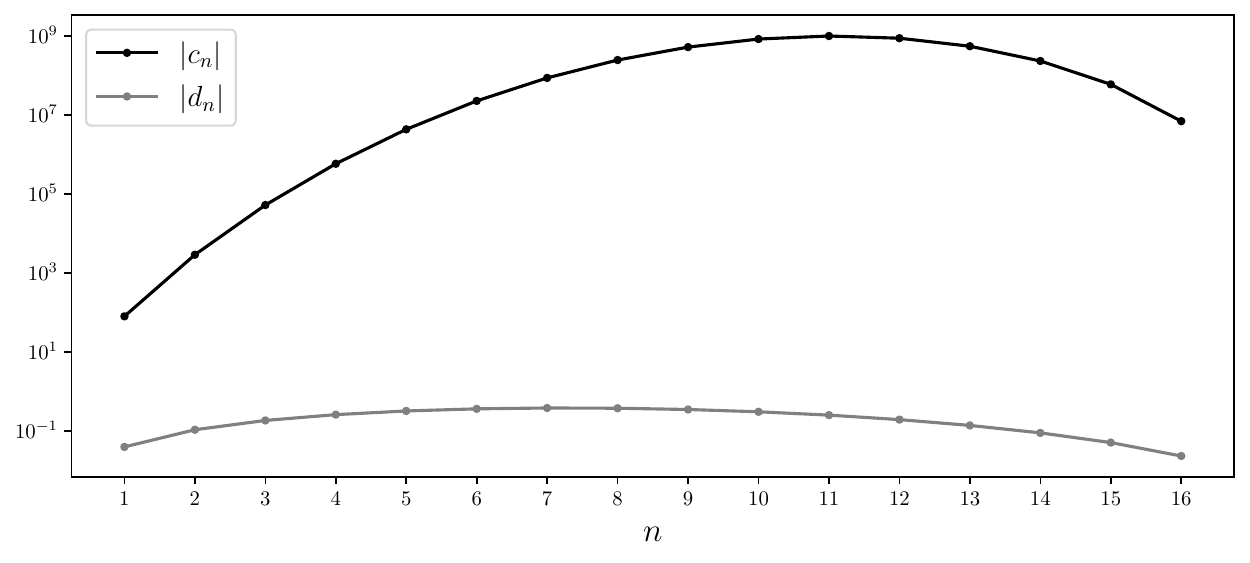}
\caption{The QNM coefficients, $c_n$, in the optimal perturbation of the de Sitter static patch shown in figure \ref{fig:dS_pick}, at the initial data surface $\tau = 0$. Also shown are the coefficients in the orthonormal basis for $W$, $d_n$. Note $\sqrt{\vec{d}^{\,*}\vec{d}} = 1$ but $\sqrt{\vec{c}^{\,*}\vec{c}} \simeq 10^{9}$. Each individual QNM decays according to the imaginary part of its eigenvalue, with the slowest-decaying coefficient $n=1$. However, the energy of the perturbation formed from this sum of 16 QNMs does not decay until approximately $\tau_* = 2.7$ due to the large cancellations that occur between non-orthogonal QNMs. }
\label{fig:dS_occupation}
\end{figure}

The magnitude of the coefficients in the eigenfunction decomposition of this optimal perturbation, $|c_n|$, are shown in figure \ref{fig:dS_occupation}. The need for large cancellations between the terms for the energy to remain at unity becomes now evident, a consequence of the non-orthogonality of QNMs.

\begin{figure}[h!]
\centering
\includegraphics[width=\columnwidth]{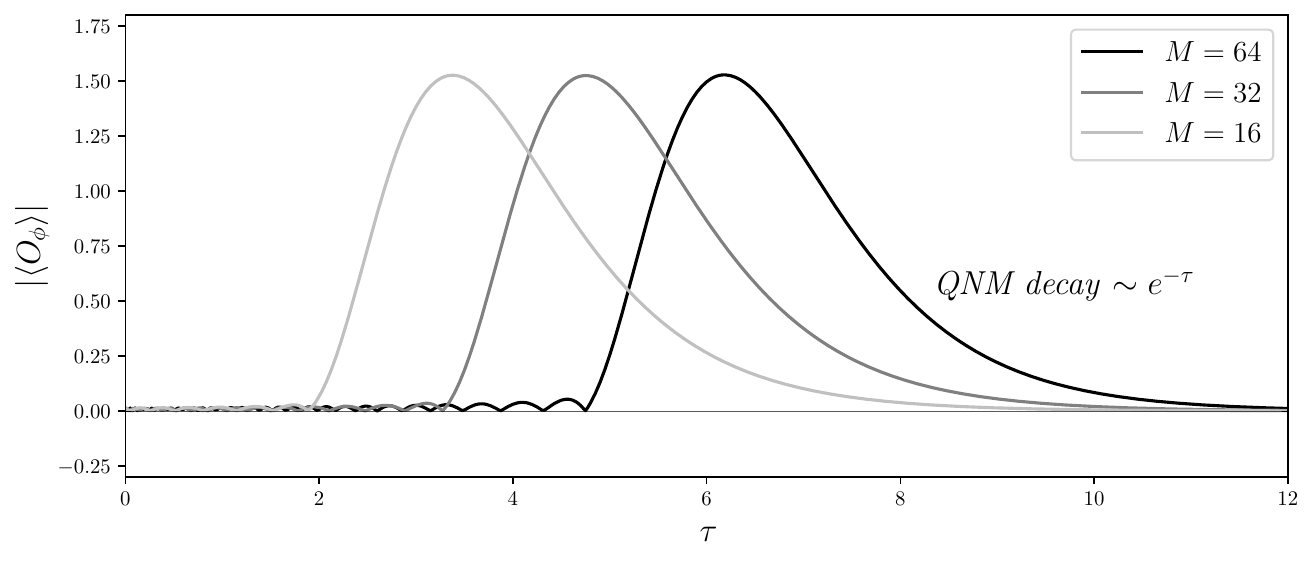}
\caption{The worldline one-point function in a static patch of de Sitter (i.e. $\left<O_\phi\right> \equiv \lim_{z\to 0} \phi$), for solutions corresponding to optimal initial data drawn from the growth curves (at $\tau_* = 2.7$ for $M=16$, $\tau_* = 4.1$ at $M=32$ and $\tau_* = 5.6$ at $M = 64$). Eventually the one-point function decays according to the longest-lived quasinormal mode, but increasing $M$ increases the transient duration and delays the onset of modal decay. The parameters used are: $m^2 = 2$, $d=3$, $l = 0$. Units are in Hubble times.}
\label{fig:dS_onepoint}
\end{figure}

Another observable of interest -- and one which is analogous to a CFT observable in the Schwarzschild-AdS example studied later -- is the worldline one-point function in the  spirit of \cite{Anninos:2011af},
\be
\left<O_\phi(\tau)\right> = \phi(\tau, 0).
\ee
This is shown in figure \ref{fig:dS_onepoint} for a set of optimal perturbations at different $M$ values. Since the optimal perturbations tend to be localised near the horizon, $\left<O_\phi(\tau)\right>$ is approximately zero until $\tau \simeq \tau_*$, when a large sudden response is exhibited. The small oscillations are reminiscent of Gibbs phenomena that result from approximating a compactly supported function by a finite number of eigenfunctions. After the peak, the decay of the perturbation is governed by the time dependence of the slowest decaying QNM, $e^{-\tau}$.

\begin{figure}[h!]
\centering
\includegraphics[width=\columnwidth]{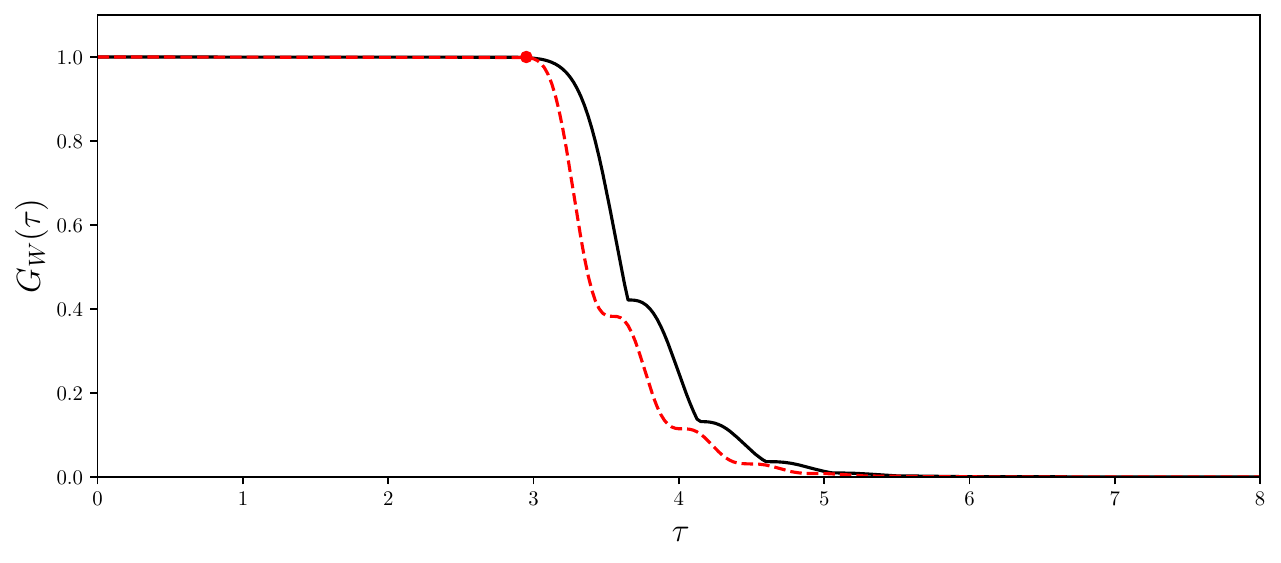}
\includegraphics[width=0.8\columnwidth]{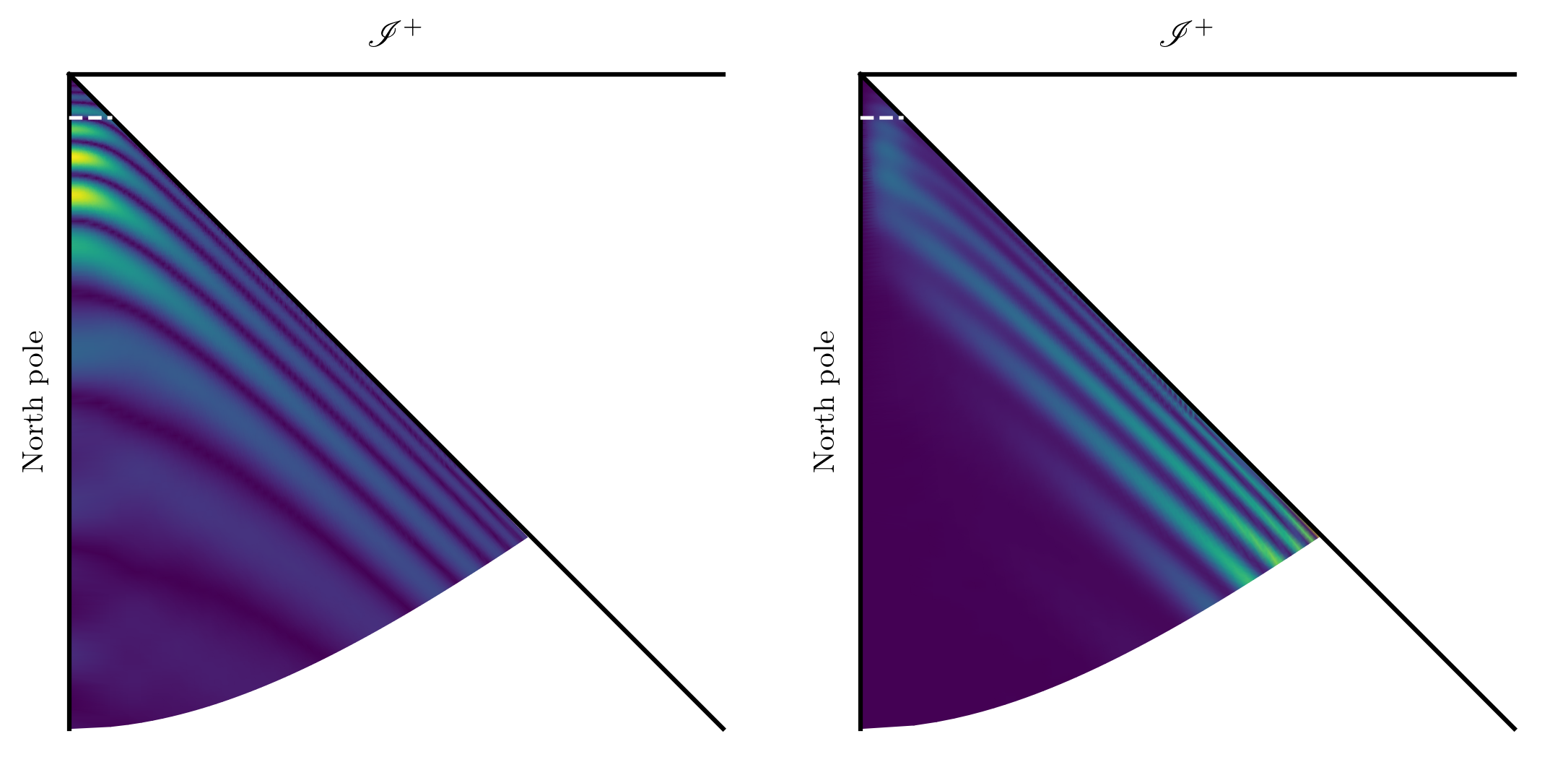}
\caption{An optimal sum of 32 QNMs of the de Sitter static patch, as in figure \ref{fig:dS_pick} but where $m^2 = 50$. This shows the evolution of the perturbation that maximises the energy at $\tau_* = 2.95$. \textbf{Upper panel:} Energy growth $G_W(\tau)$ at $M=32$ (black curve) and the energy of the optimal perturbation (red dash) that intersects the growth curve at $\tau_* = 2.95$ (red dot). Units are in Hubble times. \textbf{Lower left panel:} $|\phi|$ for this perturbation shown in the dS conformal diagram, \textbf{Lower right panel:} energy density of the perturbation. The white dashed line indicates the spatial slice $\tau_* = 2.95$, approximately where energy begins its modal decay.}
\label{fig:dS_pick_large}
\end{figure}

Finally, we construct optimal perturbations at $m^2 = 50$ (keeping $d=3$ and $l=0$). At large $m^2$, the $\omega_{nl}^\pm$ have a large real part, $\re\,\omega_{nl}^\pm \sim m$, thus the oscillation timescale is shorter than the decay timescale, here by a factor of $|\re\,\omega_{00}^\pm|/|\im\,\omega_{00}^\pm| \simeq 4.6$. The results are shown in figure \ref{fig:dS_pick_large} at $M=32$ modes. The rapid QNM oscillations are prominently visible at late times, and each oscillation peak appears at the endpoint of an energy packet propagating at the speed of light along the horizon. Thus the temporal structure of asymptotic QNM decay appears imprinted in the spatial distribution of the transient. Note that in this case the longest lived QNMs correspond to a pair of modes related under $\omega \to -\omega^*$. Correspondingly, the growth curve $G_W(\tau)$ exhibits oscillatory ringdown with `corners'. For each $\tau_*$ one can construct a perturbation with energy $G_W(\tau_*)$ at time $\tau_*$. When $\tau_*$ is in the ringdown region the perturbation is dominated by a sum of the longest-lived pair of modes. The corners are a consequence of $G_W(\tau_*)$ being the pointwise maximisation of $E(\tau_*)$, including by scanning over all possible phases between the two oscillatory modes.

\section{Schwarzschild-AdS}\label{SAdS_section}
We now study transient scalar dynamics in the exterior of planar Schwarzschild-AdS$_{d+1}$ black holes, and their dual CFT. In this section, we follow the techniques in section \ref{sec2_generalities} to construct optimal linear perturbations, which are arbitrarily long-lived.

The metric is given by \eqref{eq:lineelem_general} with planar geometry in the $d-1$ transverse directions, $d\Sigma_{d-1}^2 \equiv d\vec{x}^{\,2}$ ($\vec{x}\in \mathbb{R}^{d-1}$), and metric function
\be
f(r) = r^2-\frac{1}{r^{d-2}},
\ee
where we have fixed the AdS radius $L_{AdS}=1$, the horizon radius $r_h=1$ and we restrict to $d>1$. The black hole temperature is $T = d/(4\pi)$, and all times and frequencies in this section will be given with units $d/(4\pi T)$. We use hyperboloidal slices shown on the upper right panel of figure \ref{fig:slices} given by \eqref{eq:hyperboloidal_general} where
\bea
h(z) &=& \frac{1}{d}\log(1-z), \\
R(z) &=& \frac{1}{z}.
\eea
In these coordinates, $z=1$ corresponds to the position of the event horizon, and $z=0$ to the conformal boundary of AdS. 

The $z=0$ conformal boundary requires a choice of boundary conditions. This is one of the principal differences to the dS$_{d+1}$ case of the last section where $z=0$ is merely an origin of coordinates. Here, the boundary condition corresponds to a quantisation choice in the dual CFT and is required for a well-posed evolution of the bulk field $\Phi$. We study a scalar field dual to a dimension $\Delta$ operator in the CFT, $O_\phi$, so that $m^2=\Delta(\Delta-d)$. In particular, we restrict our attention to cases where $\Delta$ is the largest root of this equation, i.e. $2\Delta -d > 0$. The near-boundary behaviour of $\Phi$ is thus (at generic $\Delta, d$),
\be
\Phi(\tau,z,\vec{x}) = \Phi_{0}(\tau,\vec{x})z^{d-\Delta}+\ldots + \Psi_{0}(\tau,\vec{x})z^{\Delta} +\ldots,
\ee
where ellipses denote terms determined by equations of motion. The absence of sources for $O_\phi$ in the CFT becomes the Dirichlet boundary condition $\Phi_{0}(\tau,\vec{x}) = 0$. By choosing $\beta = \Delta$ in the field redefinition \eqref{eq:Phi_decomp_planewaves} so that,
\be
\phi_{\vec{k}}(\tau,z) = \phi_{0,\vec{k}}(\tau)z^{d-2\Delta}+\ldots + \psi_{0,\vec{k}}(\tau) +\ldots, \label{FefG}
\ee
the Dirichlet boundary condition $\phi_{0,\vec{k}}(\tau)=0$ becomes a regularity condition for $\phi_{\vec{k}}(\tau,z)$. With this boundary condition obeyed, the one-point function can be simply read off as \cite{Skenderis:2002wp}
\be
\left<O_\phi\right> = -(2\Delta - d) \phi_{\vec{k}}(\tau,0). \label{eq:ads1pt}
\ee
With the above considerations, we arrive at the following functions defining the energy inner product \eqref{eq:energy_innerprod} and Hamiltonian \eqref{eq:block_hamiltonian}
\bea
w(z) &=& \left(\frac{1}{\left(z^d-1\right)^2}-\frac{1}{d^2 (z-1)^2}\right) \left(1-z^d\right) z^{-d+2 \Delta +1} , \\
p(z) &=& \left(1-z^d\right) z^{-d+2 \Delta +1} , \\
q(z) &=& \left(\Delta ^2 z^d+\vec{k}^{\,2} z^2\right) z^{-d+2 \Delta -1} , \\
\gamma(z) &=& \frac{\left(1-z^d\right)}{d (z-1)} z^{-d+2 \Delta +1}.
\eea

Note that the required Dirichlet boundary conditions are satisfied by the regular, normalisable eigenfunctions of $\mathcal{H}$. We denote them by $\xi_{n}(0,z)$ with eigenvalues $\omega_n$, labelled by a single index $n$. These are the QNMs. The $\omega_n$ are poles of the retarded CFT two-point function of $O_\phi$ \cite{Birmingham:2001pj, Son:2002sd, Herzog:2002pc, vanRees:2009rw}, and typically appear in pairs related by $\omega_n \to -\omega_n^*$. 

\subsection{An analytic example for BTZ}
Before computing optimal perturbations, we examine a simple analytic example in the case of $d=2$, the BTZ black hole \cite{Banados:1992wn}, demonstrating the effects of non-normality on energy.\footnote{In this case $d\Sigma^2 = d\chi^2$ where $\chi$ is an angular coordinate and $\vec{k}$ becomes the azimuthal quantum number.} The QNMs are straightforwardly obtained \cite{Cardoso:2001hn, Birmingham:2001pj}. Adapted to hyperboloidal coordinates \eqref{eq:hyperboloidal_general}, at $\vec{k} = 0$ we find (for $\Delta > 1$)
\bea
 \phi_n(\tau, z) &=& e^{-(2n+\Delta)\tau} (z+1)^{-n-\frac{\Delta}{2}} \, _2F_1\left(-n,-n;1-2 n-\Delta;1-z^2\right),\\
\omega_n &=& -i (\Delta +2 n),
\eea
for $n= 0, 1, \ldots$, as the eigenfunctions and associated regular eigenvalues of $\mathcal{H}$ with no source term at the boundary, with arbitrary normalisation. 

Let us focus on $\Delta = 2$ ($m^2 = 0$) perturbations for simplicity. We first form our perturbation from a sum of the $n=0$ and $n=1$ modes,
\be
\phi(\tau, z) = a_1 \frac{e^{-2\tau}}{1+z} + a_2 \frac{e^{-4\tau} (2+z^2)}{3(1+z)^2}, \label{eq:BTZ2mode}
\ee
whose total energy is 
\be
E(\tau) = \frac{1}{4}|a_1|^2 e^{-4\tau} + \frac{1}{6}(a_1^*a_2 + a_1 a_2^*)e^{-6\tau} + \frac{1}{8} |a_2|^2 e^{-8\tau}.
\ee
Note the appearance of the cross-term -- without this term the slowest energy decay is achieved by setting $a_2 = 0$. In the neighbourhood of $\tau = 0$ one has
\be
\frac{E(\tau)}{E(0)} = 1 - 24\frac{|a_1+a_2|^2}{6 |a_1|^2 + 4(a_1^*a_2 + a_1 a_2^*) + 3 |a_2|^2}\tau + O(\tau)^2
\ee
so that a choice $a_2 = -a_1$ removes the $O(\tau)$ term (and also removes the $O(\tau)^2$ term), delaying the onset of energy decay. As in the analogous de Sitter example \eqref{examplephisol}, including more modes in the sum \eqref{eq:BTZ2mode} allows for further terms in this Taylor series to be removed. For $M$ modes, the following choice of coefficients achieves this goal,
\be
\phi(\tau, z) = \sum_{n=0}^{M-1} {M-1 \choose n} \frac{(-1)^n 2^{n+1}}{n+1}\sqrt{M(2M-1)} \,\phi_n(\tau,z), \label{sadsexample}
\ee
whose energy in the neighbourhood of $\tau = 0$ is given by
\be
E(\tau) = 1 - 4^M M\tau^{2M - 1} + \ldots \,.
\ee
Whilst directly performing the sum in \eqref{sadsexample} does not obviously yield a useful closed-form expression for $\phi(\tau, z)$, the flux is given by a boundary term \eqref{eq:FluxCC} which can easily be evaluated. Then, energy follows by integrating with respect to $\tau$,
\bea
E(\tau) &=& 1 - (1-e^{-2\tau})^{2M-1}\left(1 + (2M-1)e^{-2\tau}\right),\\
\mathcal{F}(\tau) &=& 4 M(2M-1) e^{-4\tau}(1-e^{-2\tau})^{2M-2}.
\eea
As is by now familiar, $E(\tau)$ displays a transient plateau whose duration logarithmically increases with $M$; the peak flux through the future horizon occurs at $\tau = \frac{1}{2}\log M$. The associated CFT one-point function \eqref{eq:ads1pt} can also be obtained,
\be
\left<O_\phi\right> = -4 e^{-2\tau}\sqrt{M(2M-1)}\;{}_2F_1\left(2,1-M;\frac{3}{2}; \frac{e^{-2\tau}}{2}\right),
\ee
displaying a pulse, when the energy packet reflects off the boundary, at a time that grows logarithmically with $M$.

This example proves, at least for $\vec{k}=0$, $\Delta = 2$, that for any $\tau \geq 0$ one can construct $W$ containing perturbations such that $E(\tau)$ is arbitrarily close to one (by using a suitably large $M$). Through \eqref{GWboundsG} this implies $G(\tau) = 1$ in this case.
One can also verify that the lifetime is infinite for an infinite sum,
\be
\lim_{M \to \infty} E(\tau) = 1,\qquad \lim_{M \to \infty} {\cal F}(\tau)  = 0\qquad (\tau \geq 0),
\ee
but just as in the de Sitter example, $\partial_z\phi(\tau, z)|_{z=1}$ diverges with $M$, and hence only finite $M$ should be considered. Thus we have constructed finite sums of BTZ QNMs with an arbitrarily long but finite lifetime.

Finally, we note that a similar approach can be taken to constructing solutions by removing terms in the $\tau$ Taylor series of $\left<O_\phi(\tau)\right>$, rather than $E(\tau)$. We find a solution of this type (with arbitrary normalisation) given by,
\bea
\phi(\tau, z) &=& \sum_{n=0}^{M-1} {2n+1 \choose n+1} \frac{(-1)^n \Gamma(M+1)}{\Gamma(M-n)\Gamma(n+2)} \,\phi_n(\tau,z),\\
\left<O_\phi\right> &=& -2M e^{-2\tau}(1-e^{-2\tau})^{M-1}.
\eea

\subsection{Energy growth and optimal perturbations}\label{sec:ads_numerics}
We now turn to computing optimal perturbations for $d \geq 2$ in the Schwarzschild-AdS$_{d+1}$ black brane, proceeding numerically using spectral methods.

We start by computing the QNMs eigenfunctions $\xi_{n}(0,z)$ and eigenvalues $\omega_n$ labelled by $n = 1, 2, \ldots$. We discretise $\mathcal{H}$ into a $2(N+1)\times2(N+1)$ matrix, $H$, whose numerical eigenvectors correspond to the QNMs. Note that in this process $H$ may contain spurious unphysical eigenvectors as a result of the numerical approximation, that we discard. Furthermore, care must be taken when constructing the rows of $H$ corresponding to the conformal boundary $z=0$, depending on the particular approach taken. For $\p_\tau \phi$ at $z=0$ we impose the condition developed in the near-boundary expansion, $(d\p_z\phi - \p_\tau \phi)\big|_{z=0}=0$, by replacing the $(N+2)^{\text{th}}$ row, leading to a generalised eigenvalue problem $H \vec{\xi} = \omega B \vec{\xi}$, with $B$ the $2(N+1)\times2(N+1)$ identity matrix with the $(N+2)^{\text{th}}$ row replaced by zeroes. Finally, in similar fashion, the energy inner product is discretised into $\left< \vec{\xi}_1, \vec{\xi}_2\right> = \vec{\xi}_{1}^{\,*} \,G\, \vec{\xi}_2 = (F\vec{\xi}_{1})^*(F\vec{\xi}_{2})$, where $G$ is a $2(N+1)\times2(N+1)$ matrix and $F$ its Cholesky decomposition. For further details see appendix \ref{sec:spectral_methods}.

We now turn to the recipe in section \ref{sec2_generalities}.
With $\xi_{n}(0,z)$, $\omega_n$ computed numerically, we take the $M$ QNMs closest to the real $\omega$ axis forming the subspace $W$. First, we unit-normalise the eigenvectors in the energy norm, $\vec{\xi}_n$, from which we build the $2(N+1)\times M$ matrix of eigenvectors $V_W=\left( \vec{\xi}_1 \; \vec{\xi}_2 \; \ldots \; \vec{\xi}_M \right)$. Following \cite{Reddy93}, we then orthogonalise these eigenvectors with respect to the energy inner product using the QR decomposition of the matrix $FV_W$, giving us $(FQ_W)$ and $U_W$ such that\footnote{In Mathematica, $FQ_W$ and $U_W$ can be obtained using the function $\verb|{q,r} = QRDecomposition[|FV_W\verb|]|$ where $FQ_W = \verb|ConjugateTranspose[q]|$ and $U_W = \verb|r|$.}
\be
FV_W = (FQ_W) U_W ,
\ee
where the columns of $Q_W$ now comprise the new orthonormal basis $Q_W = \left( \vec{\psi}_1 \; \vec{\psi}_2 \; \ldots \; \vec{\psi}_M \right)$ for $W$ in the energy norm, i.e. $\left<\vec{\psi}_i , \vec{\psi}_j \right> = (F\vec{\psi}_{i})^*(F\vec{\psi}_{j})=\delta_{ij}$. The $U_W$ matrix is the $M\times M$ transformation matrix mapping coefficients in a QNM decomposition to those in an orthonormal decomposition $\vec{d}=U_W \vec{c}$, as in \eqref{eq:xiexpansion_M_orthbasis}. Finally, $D_W = \text{diag}(\omega_1,\omega_2,\ldots,\omega_M)$, and thus we have all the ingredients to compute $H_W$ \eqref{HWdef}, hence the growth curve $G_W(\tau)$ \eqref{eq:GW_2norm}.

\begin{figure}[h!]
\centering
\includegraphics[width=\columnwidth]{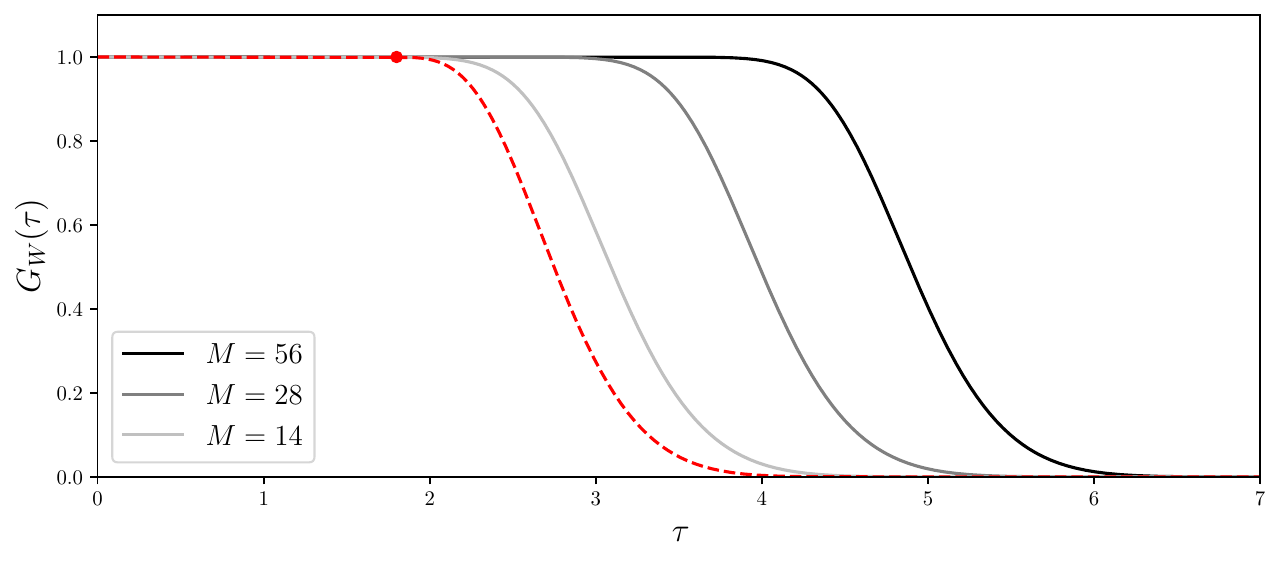}
\includegraphics[width=0.8\columnwidth]{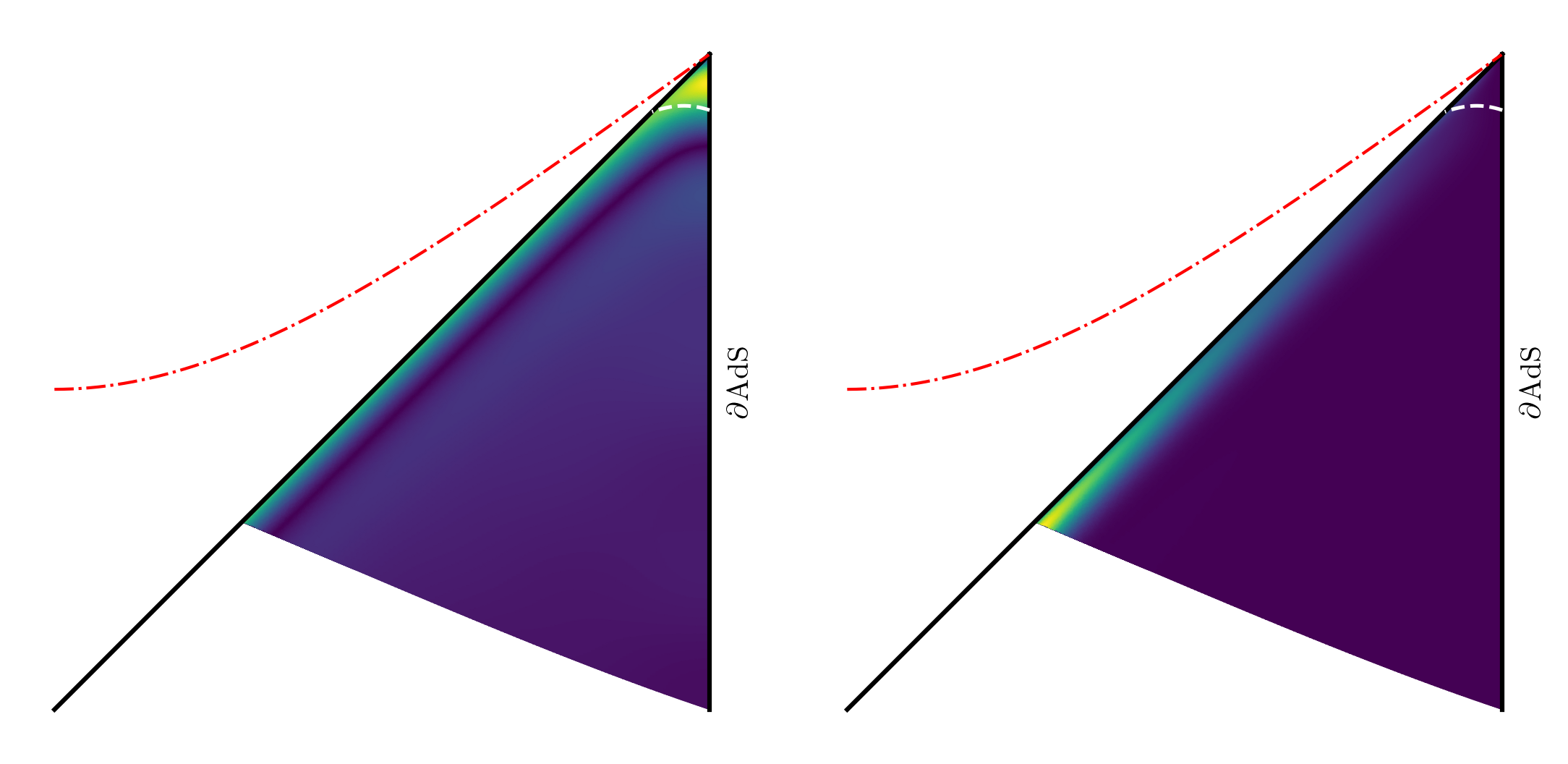}
\caption{An optimal sum of QNMs in the exterior of planar Schwarzschild-AdS. Here $M=14$ scalar field QNMs maximise the energy at $\tau_*=1.8$. Parameters used: $\Delta=2$, $d=3$, $\Vec{k}=0$. \textbf{Upper panel:} Energy growth curves $G_W(\tau)$ for various $M$ (solid curves), with the energy of the optimal perturbation for $\tau_*=1.8$ at $M=14$ (red dash). \textbf{Lower left panel:} Colour indicates the values of $|\phi|$ for the optimal perturbation, shown in the Schwarzschild-AdS conformal diagram. \textbf{Lower right panel:} Colour indicates the energy density of the optimal perturbation in the Schwarzschild-AdS conformal diagram. The energy is localised near the event horizon and gets closer to the horizon for larger values of $M$. The dashed white slice is the spatial slice at $\tau_*=1.8$, approximately where the energy begins its modal decay.}
\label{fig:SAdS_pick_14}
\end{figure}

The results are as follows.\footnote{We use $N=200$ and $200$ digits of precision.} The upper panel in figure \ref{fig:SAdS_pick_14} shows $G_W(\tau)$ for the solutions with $\Delta=2, d=3, \vec{k}=0$, in the subspace of $M=14, 28, 56$ QNMs, exhibiting the transient plateau. On top of it, we plot the energy for the $M=14$ optimal perturbation picked at $\tau_* = 1.8$, exhibiting transient dynamics that tracks the plateau. The spatial distribution of the field and its energy density (the integrand in \eqref{eq:energy_innerprod}) are displayed in the conformal diagrams of figure \ref{fig:SAdS_pick_14}. They show a perturbation localised near and propagating along the horizon. Just as in the de Sitter case of section \ref{dS_section}, increasing $M$ leads to a logarithmically increasing transient plateau duration, and a perturbation which gets ever-closer to the horizon.

\begin{figure}[h!]
\centering
\includegraphics[width=\columnwidth]{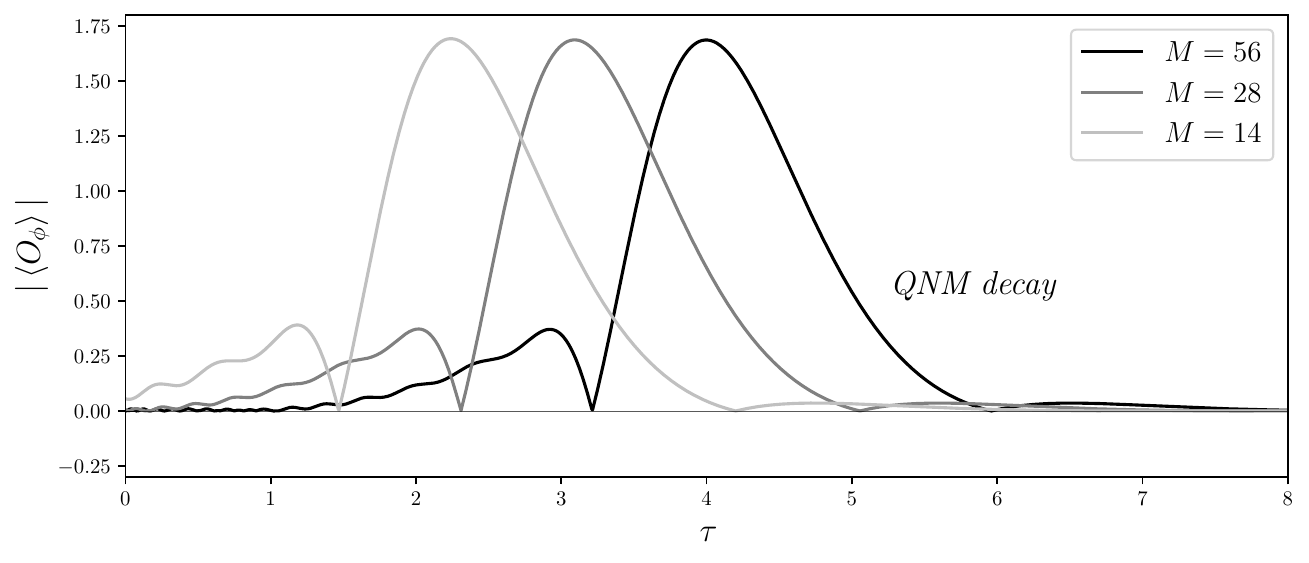}
\caption{The CFT one-point function of the operator dual to the scalar field $\Phi$, for states prepared by optimal initial data drawn from the growth curves ($\tau_* = 1.8$ at $M=14$, $\tau_* = 2.6$ at $M=28$ and $\tau_* = 3.5$ at $M=56$). During the transient period $0 \leq \tau \lesssim \tau_*$, the one-point function exhibits a moderate response, after which a large sudden response is seen at $\tau \simeq \tau_*$. Eventually, it decays according to the pair of longest-lived quasinormal modes, but increasing $M$ increases the transient duration and delays the onset of modal decay. The parameters used are: $\Delta=2$, $d=3$, $\Vec{k}=0$.}
\label{fig:SAdS_onepoint_abs_142856}
\end{figure}

Perturbations with nonzero transverse momenta $\vec{k}$ display similar behaviour, but the transient plateau is shorter. Since different $\vec{k}$ modes are orthogonal, the $\vec{k} = 0$ modes we have constructed here therefore provide the optimal contributions, even when allowing for nontrivial profiles in the transverse directions.

Turning to the CFT dual, we compute the one-point function of $O_\phi$ in the states prepared by choosing initial conditions corresponding to the optimal perturbations, using \eqref{eq:ads1pt}. The one-point function is shown for three optimal perturbations with different $M$ and $\tau_*$ in figure \ref{fig:SAdS_onepoint_abs_142856}. We observe an initial period where $\left<O_\phi(\tau)\right> \simeq 0$, after which it exhibits large sudden response starting at $\tau \simeq \tau_*$, when the near-horizon energy packet hits the boundary, followed by QNM decay. By increasing $M$, one may delay the appearance of the pulse, thus describing a state with $\left<O_\phi(\tau)\right>\simeq 0$ for an arbitrarily long period of time. We discuss the relation to the thermalisation of holographic plasmas in section \ref{sec:discussion}.

\section{Schwarzschild}\label{flat_section}
In this section we construct optimal perturbations of asymptotically flat, $3+1$-dimensional Schwarzschild black holes. The geometry \eqref{eq:lineelem_general} with $d=3$ has a transverse 2-sphere $d\Sigma_2^2 = d\Omega_2^2$, and the metric function is given by
\be
f(r) = 1-\frac{1}{r}, 
\ee
where, without loss of generality, we work in units where the event horizon radius $r_H=1$. We follow closely the treatment of perturbations given in \cite{Jaramillo:2020tuu} where hyperboloidal slices \eqref{eq:hyperboloidal_general} are prescribed (lightly adapted to our conventions)\footnote{In particular, compared to \cite{Jaramillo:2020tuu}, our stable frequencies sit in the lower-half $\omega$ plane, while $\tau = 2\tau_\text{there}$ which changes various normalisations. This includes frequencies $\omega = \frac{1}{2}\omega_\text{there}^*$ and the energy inner-product through $w = 4\times \frac{1}{2} w_\text{there}$, $p = \frac{1}{2}p_\text{there}$, $q = \frac{1}{2}q_\text{there}$. These factors come from the definition of the energy (involving the contraction with $(\partial_\tau)^\mu$) and the time derivatives that $w$ multiplies. As compared to e.g. \cite{Leaver:1985ax}, our $\omega = \omega_\text{there}^*$ for $l=2$, $s=2$, and we confirm agreement with the QNM frequencies quoted there. } as follows
\bea
h(z) &=& \log(z) + \log(1-z) - 1/z, \\
R(z) &=& \frac{1}{z}.
\eea
Now $z=1$ corresponds to the future event horizon while $z=0$ corresponds to future null infinity, as depicted in figure \ref{fig:slices}. When $m^2=0$ and with $\beta = 1$ in \eqref{eq:Phi_decomp_harmonics}, the scalar field perturbation $\Phi$ corresponds with the $s=0$ scalar appearing in the Regge-Wheeler equations. To incorporate a larger set of physically interesting perturbations, we now extend our analysis from $\Phi$ to the full set of Regge-Wheeler-Zerilli perturbations (see for example \cite{Kokkotas:1999bd, Berti:2009kk, Jaramillo:2020tuu}), while utilising the same form of the energy inner-product \eqref{eq:energy_innerprod}, closely following \cite{Jaramillo:2020tuu}. We have
\bea
w(z) &=& 4(1+z) , \\
p(z) &=& z^2(1-z) , \\
\gamma(z) &=& 1-2z^2,
\eea
and for axial/Regge-Wheeler perturbations with spin $s$, we have,
\be
q(z) = l(l+1)+(1-s^2)z \,,
\ee
while for polar/Zerilli perturbations,
\be
q(z) = \frac{1}{3} (l-1) (l+2) \left(\frac{2 (l-1) (l+2) \left(l^2+l+1\right)}{\left(l^2+l+3 z-2\right)^2}+1\right)+z \,.
\ee
The property that $\gamma(0) = 1$ and $\gamma(1) = -1$ for both of these perturbation types is what leads to the two sign-definite contributions to the flux in \eqref{eq:FluxFlat}.

\begin{figure}[h!]
\centering
\includegraphics[width=0.6\columnwidth]{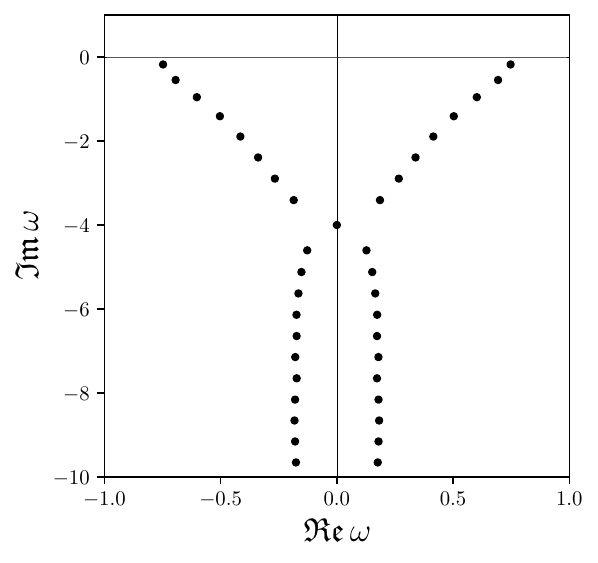}
\caption{The $l=2$ spectrum of gravitational perturbations of Schwarzschild that we use to form the subspace $W$. Here $M=39$ modes are shown. Units are given by $r_H = 1$. \label{fig:flat_modes}}
\end{figure}

With all the relevant functions defined, the numerical analysis closely parallels that presented in section \ref{sec:ads_numerics}. We direct the reader there for more details, including the construction of operators $U_W$ and $H_W$.\footnote{Note that, in this case, one does not need to replace any rows in the matrix eigenvalue problem.} We focus here on perturbations formed in the Hilbert space $W$, comprised of QNMs in the $s=2$, $l=2$ Regge-Wheeler spectrum. The $s=2$, $l=2$ Regge-Wheeler and $l=2$ Zerilli perturbations are isospectral. We note that there are a large number of spurious unphysical eigenfunctions of $\mathcal{H}$ in this case, whose eigenvalues are distributed along the imaginary $\omega$ axis. There are two ways to see that they are unphysical: $\phi$ diverges at future null infinity ($z=0$) with increasing numerical resolution, and $\omega$ does not converge with increasing numerical resolution. They are usually attributed to the inability of spectral methods to resolve a branch point singularity in the retarded Green's function at $\omega = 0$. We therefore do not include them in $W$. Here $W$ is formed from the first $M$ QNMs, in order of increasing distance from the real axis, taking care to exclude spurious modes while keeping the mode at $\omega \simeq -4i$. An example of these selections is shown in figure \ref{fig:flat_modes}.

\begin{figure}[h!]
\centering
\includegraphics[width=0.9\columnwidth]{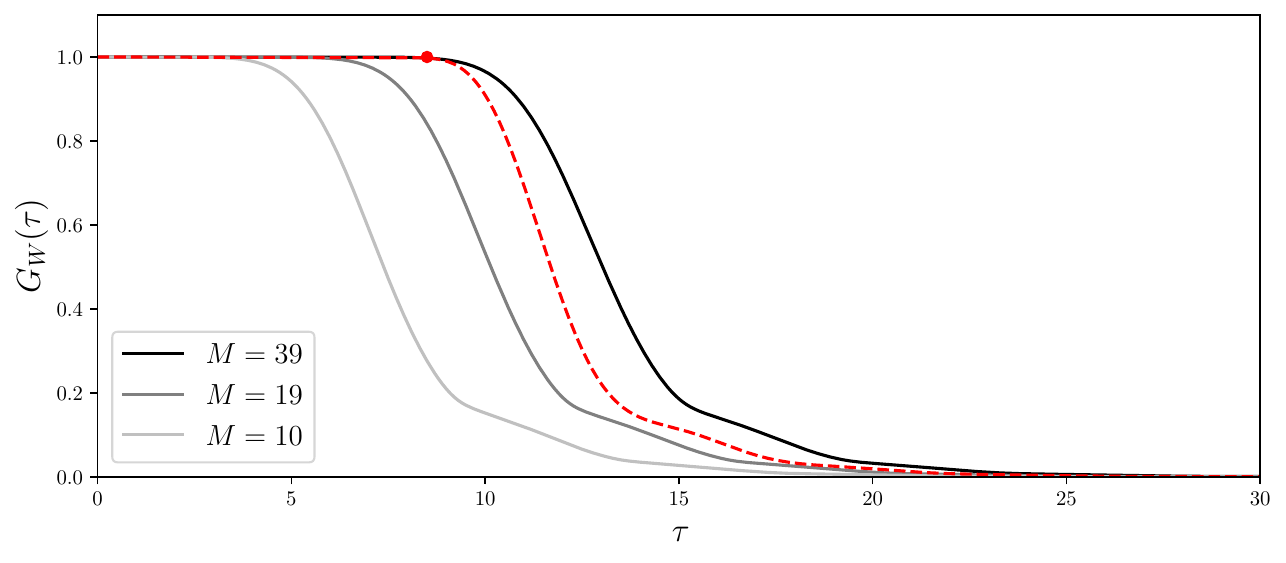}
\includegraphics[width=\columnwidth]{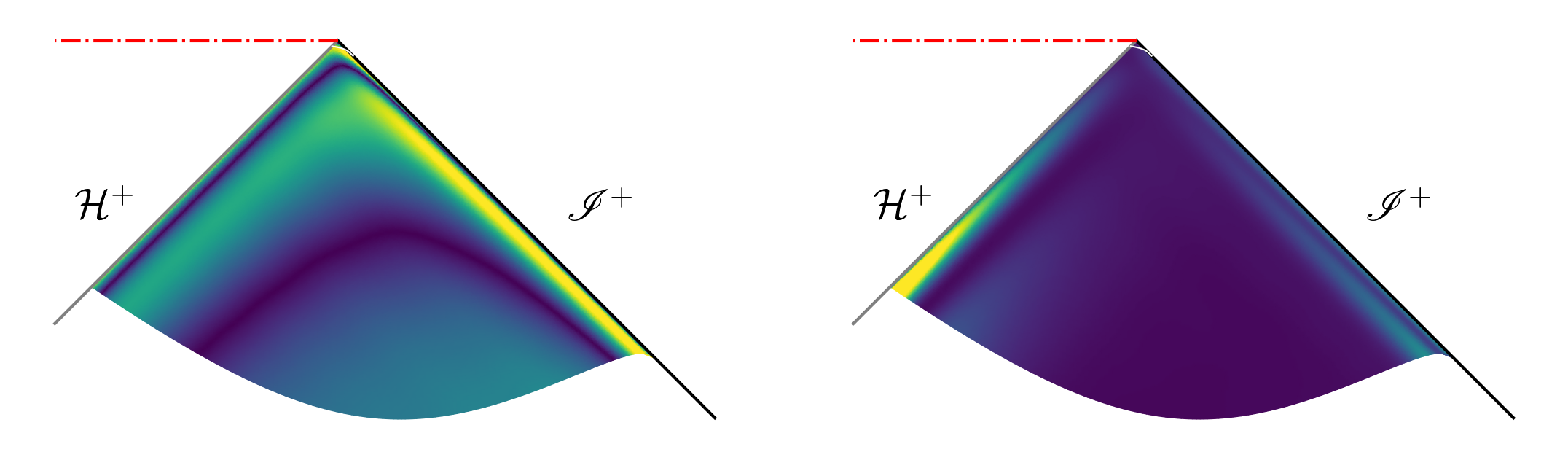}
\caption{\textbf{Upper panel:} The energy growth curve for $l=2$ gravitational perturbations of the Schwarzschild spacetime for various $M=\dim(W)$ (solid curves), and the energy of an optimal sum of $M=39$ QNMs (dashed red curve) that maximise the energy at $\tau_* = 8.5$ (red dot). The spectrum of $W$ is as in figure \ref{fig:flat_modes}.  \textbf{Lower left panel:} Colour indicates the values of $|\phi|$ for the optimal perturbation in a portion of the Schwarzschild conformal diagram. \textbf{Lower right panel:} Colour indicates the energy density of the optimal perturbation in the conformal diagram. The energy is initially localised near the future event horizon and future null infinity, then propagates along them. The packet disperses and then eventually begins to leave the domain at $\tau \simeq \tau_*$ (barely visible white slice near $i^+$).
\label{fig:flat_pick}}
\end{figure}

The results are shown in figure \ref{fig:flat_pick} for a choice of $M=10,19,39$ QNMs.\footnote{We use $N=299$ and $200$ digits of precision.} The selection of modes is shown in figure \ref{fig:flat_modes}, and the spectrum is in agreement with \cite{Leaver:1985ax}. $G_W(\tau)$ displays a transient plateau, which increases in duration with $M$ at a logarithmic rate, as in all other examples studied in this paper. Selecting a time $\tau_*$ on the growth curve, we construct the optimal transient perturbation at $M=39$ which obeys $E(\tau_*) = G_W(\tau_*)$, through the singular value decomposition of $e^{-i H_W\tau}$.\footnote{For the avoidance of doubt, we confirm that the energy of either the real part or imaginary part of the optimal perturbation, as a function of $\tau$, is the same as the total energy $E(\tau)$, once normalised to one at $\tau = 0$.} Its distribution in spacetime is shown in the conformal diagram, displaying energy packets that propagate along the future horizon and along future null infinity.

\section{Relation to pseudospectra}\label{sec:pseudospectra}
There is an intimate mathematical connection between transient dynamics for a non-normal Hamiltonian, $\mathcal{H}$, and its pseudospectrum \cite{TrefethenEmbree2005}. Heuristically, the pseudospectrum $\sigma_\epsilon(\mathcal{H})$ is a one-parameter family that enlarges the spectrum $\sigma(\mathcal{H})$ to include frequencies $\omega$ which are `almost eigenvalues'. We may either view the `almost eigenvalues' as those $\omega$ at which the system displays a suitably large response, according to the definition\footnote{Here we require $\mathcal{H} \in \mathscr{C}(X)$, the set of closed operators on a complex Banach space $X$.}
\be
\sigma_{\epsilon}(\mathcal{H}) = \{\omega \in \mathbb{C} : \|R(\omega;\mathcal{H})\| \geq \epsilon^{-1}\} \label{eq:pseudo_def_resolvent}
\ee
where we introduce the resolvent operator $R(\omega;\mathcal{H}) \equiv (\omega - \mathcal{H})^{-1}$, or, we view them as eigenvalues of a system which is suitably nearby, according to the definition
\be
\sigma_{\epsilon}(\mathcal{H}) =\{\omega \in \mathbb{C} : \omega \in \sigma(\mathcal{H} + \delta\mathcal{H}),\,\|\delta\mathcal{H}\| \leq \epsilon\}. \label{eq:pseudo_def_perts}
\ee
The two definitions \eqref{eq:pseudo_def_resolvent} and \eqref{eq:pseudo_def_perts} are equivalent \cite{TrefethenEmbree2005}.
Evidently, the meaning of `almost' is determined by the parameter $\epsilon$ compared to distances evaluated in the choice of norm, $\|\cdot\|$, which in principle could be taken to be large.

The first connection we wish to highlight is captured by (half of) Theorem 15.4 in \cite{TrefethenEmbree2005} (see also the Kreiss Matrix theorem, Theorem 18.5 there). Applied to the present context, it states
\be
\mathcal{K}^2(\mathcal{H}) \leq  \sup_{\tau \geq 0}G(\tau) , \label{KMT}
\ee
\sloppy where $\mathcal{K}(\mathcal{H})$ is the Kreiss constant of $\mathcal{H}$, defined as $\mathcal{K}(\mathcal{H}) = \sup_{\im \,\omega >0}(\im\,\omega) \|R(\omega;\mathcal{H})\|_E$. Combining the upper bound \eqref{Gbounds} with \eqref{KMT} and noting that $(\im\,\omega) \|R(\omega;\mathcal{H})\|_E\leq \sup_{\im \,\omega >0}(\im\,\omega) \|R(\omega;\mathcal{H})\|_E$ for all $\omega\in\mathbb{C}$ with $\im\,\omega > 0$, we have that
\be
\|R(\omega;\mathcal{H})\|_E \leq \frac{1}{\im{\,\omega}}\quad \forall \omega \in \mathbb{C} \; \text{s.t.}\;\im\,\omega > 0. \label{ResolventBoundedness}
\ee
Thus the statement of no energy growth becomes a bound on pseudospectrum contours in the upper half $\omega$ plane in the energy norm, i.e. 
\be
\im\,\omega \leq \epsilon  \quad \forall  \omega \in \sigma_\epsilon(\mathcal{H}) \;\text{s.t.}\; \im\,\omega > 0. \label{PSproperty}
\ee

The second connection relates the protrusion of pseudospectra into the upper-half $\omega$ plane with $G'(0)$ (see Theorem 17.4 of \cite{TrefethenEmbree2005}),
\be
G'(0) = 2\,\sup_{\epsilon > 0} \left(\sup_{\omega \in \sigma_{\epsilon}(\mathcal{H})} \im{\,\omega} - \epsilon\right).\label{Gprimezero}
\ee
From our computations of $G(\tau)$ in previous sections, we have seen that $G'(0) = 0$, and thus this will be reflected in the properties of the pseudospectrum.

We now turn to computing $\sigma_\epsilon(\mathcal{H})$ in the energy norm for perturbations of the de Sitter static patch as in section \ref{dS_section}. We will confirm the expected behaviour \eqref{ResolventBoundedness}, \eqref{PSproperty}, \eqref{Gprimezero}, and discuss the associated consequences for the numerical convergence of $\|R(\omega;\mathcal{H})\|_E$ with $N$ in the upper-half $\omega$ plane. The pseudospectrum for scalar QNMs in the static patch of dS$_2$ with $m^2=1$, more commonly known as the Pöschl-Teller model, was computed in \cite{Jaramillo:2020tuu}. Here we provide the techniques required to generalise to any $d$ in dS$_{d+1}$ and any value of $m^2$, though for concreteness we focus on the case $d=3$, $m^2 = 2$ in line with the first example considered in section \ref{sec:dS_energygrowth}.

In order to compute $\sigma_\epsilon(\mathcal{H})$, we use spectral methods, which approximate functions by truncated expansions of Chebyshev polynomials, working with $N+1$ collocation points in the $z$ direction. Thus, fields $\xi = (\phi, \partial_\tau\phi)^T$ become $2(N+1)$-dimensional vectors, $\vec{\xi}$, while the differential operator $\mathcal{H}$ becomes a $2(N+1) \times 2(N+1)$ matrix, $H$. Similarly, the energy inner product can be computed as \eqref{eq:energy_innerprod} $\left< \xi_1, \xi_2\right> = \vec{\xi}_{1}^{\,*} \,G\, \vec{\xi}_2$, where $G$ is a $2(N+1) \times 2(N+1)$ matrix. Finally, the discretised form of \eqref{eq:pseudo_def_resolvent} in the energy norm becomes \cite{TrefethenEmbree2005}
\be
\sigma_{\epsilon}(H) = \{\omega \in \mathbb{C} : s_{\text{min}}(\omega - FHF^{-1}) \leq \epsilon \},
\ee
where $s_{\text{min}}(B)$ denotes the minimum singular value of the matrix $B$, and $F$ is the Cholesky decomposition of $G$, i.e. $G=F^*F$. See appendix \ref{sec:spectral_methods} for a more detailed discussion. 

\begin{figure}[h!]
\centering
\includegraphics[width=0.8\columnwidth]{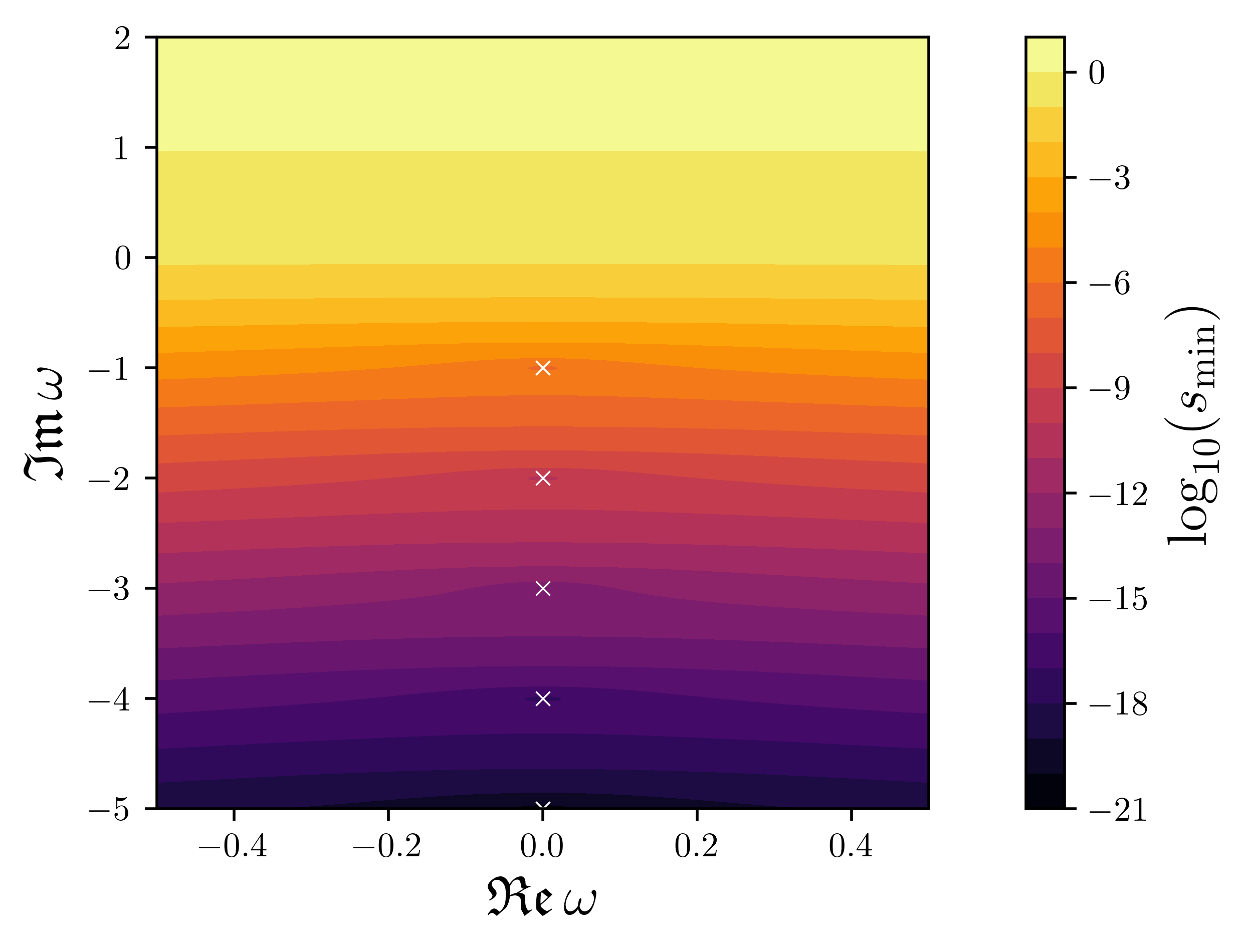}
\caption{Contours of the pseudospectrum $\sigma_\epsilon(\mathcal{H})$ in the energy norm, governing linear scalar perturbations of the de Sitter static patch. The values of the QNM frequencies are shown with white crosses, corresponding to $\omega = -i,-2i,-3i,\ldots$, and colour indicates $\log_{10}(s_{\min}(\omega-FHF^{-1}))$. The $\epsilon$-pseudospectra, $\sigma_{\epsilon}(\mathcal{H})$, form nested open sets delimited by the contours $s_{\min}(\omega-FHF^{-1}) = \epsilon$, indicating the presence of a spectral instability. Parameters used: $m^2 = 2$, $d=3$, $l=0$, with numerical resolution $N=100$ and $80$ digits of precision.}
\label{fig:dS_pseudo}
\end{figure}

Figure \ref{fig:dS_pseudo} shows the contours of the pseudospectrum of scalar QNMs of the dS$_4$ static patch with $m^2=2$, $l=0$, at $N=100$ and using $80$ digits of precision.\footnote{We choose these parameters to match those of figures \ref{fig:dS_pick} and \ref{fig:dS_occupation}.}  This indicates a spectral instability since the $\epsilon$-pseudospectrum contours form open sets.\footnote{See, however, the comments below regarding convergence with $N$ in the lower-half plane.} By \eqref{eq:pseudo_def_perts}, open pseudospectra sets imply that there exist perturbations to $\mathcal{H}$ under which QNMs move distances that are much greater than the size of the perturbation. This is a different, interesting consequence of non-normality which we do not explore further.

\begin{figure}[h!]
\centering
\includegraphics[width=0.6\columnwidth]{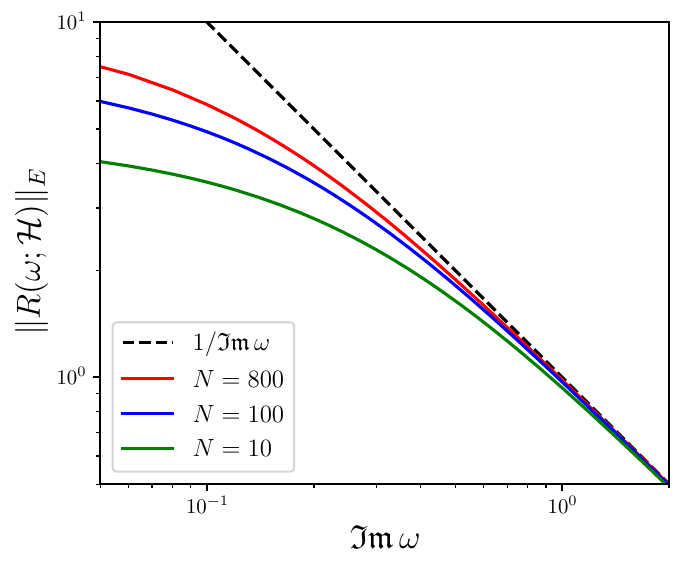}
\caption{Continuum convergence test for $\|R(\omega;\mathcal{H})\|_E$ governing linear scalar perturbations to the static patch of de Sitter. Each solid line corresponds to a different numerical resolution $N=10,100,800$. The lines display agreement with the no-growth upper-bound \eqref{ResolventBoundedness}, indicated by the dashed line. Additionally, $\|R(\omega;\mathcal{H})\|_E$ shows evidence of converging to $1/\im\,\omega$ at every point in the upper half plane as $N\to\infty$. Parameters used: $m^2 = 2$, $d=3$, $l=0$. The precision used is $80$ digits.}
\label{fig:dS_conv}
\end{figure}

Finally, we confirm the no-growth bound \eqref{ResolventBoundedness} in figure \ref{fig:dS_conv} where we sample an interval along the positive imaginary $\omega$ axis.\footnote{The results for an interval with a constant non-zero real part are qualitatively the same, but with a slower convergence farther away from the imaginary axis.} We see that \eqref{ResolventBoundedness} is obeyed, and furthermore the bound appears to saturate as $N$ is increased, consistent with $G'(0)=0$ according to \eqref{Gprimezero}. The bound \eqref{ResolventBoundedness} is independent of the system under consideration, linked only to the non-growth of the maximum linear response to initial data in a given norm. We anticipate the pseudospectrum will converge with $N$ in the upper half of the complex plane for any gravitational system with these characteristics.

In the lower-half plane, $\|R(\omega;\mathcal{H})\|_E$ does not appear to converge with $N$. This is perhaps not surprising since $R$ can be understood as the frequency-space Green's function in the presence of an external source, $s(\omega)$, added to the right-hand side of \eqref{eq:block_hamiltonian}, viz.
\be
\xi(\omega) = R(\omega, \mathcal{H})s(\omega).
\ee
A diverging $\|R(\omega;\mathcal{H})\|_E = \sup_{s}\frac{\|\xi(\omega)\|_E}{\|s\|_E}$ with increasing $N$ then indicates that, with such a source, one can elicit a larger response from the system by utilising shorter and shorter wavelength perturbations in an unbounded fashion. We confirm this interpretation by computing the `optimal source', $s(\omega)$, using the singular value decomposition of $R(\omega, \mathcal{H})$ -- it is peaked at the horizon with oscillations on the scale of the grid spacing, $1/N$. If this were a concern one could choose to penalise higher spatial gradients in a modified norm (see for instance \cite{Warnick:2013hba, Boyanov:2023qqf}), however, given the arbitrariness of $s$ as a bulk source term, this does not seem to be physically well-motivated. This is in stark contrast to the situation in the upper-half plane where $\|R(\omega;\mathcal{H})\|_E$ is necessarily bounded due to lack of energy growth.

\section{Discussion}
\label{sec:discussion}

In this work we have examined the consequences of non-normality in black hole perturbation theory. Importing techniques from the hydrodynamics literature, we constructed optimal sums of QNMs that are arbitrarily long-lived, corresponding to initial conditions in which a packet of energy is localised along the future horizon. Individually, each QNM decays exponentially, but together, decay can be postponed indefinitely due their overlap in the relevant inner product, with a maximum lifetime that grows logarithmically with the number of modes. Our results appear to be universal, with qualitatively similar results found in AdS black holes, asymptotically flat black holes, and cosmological horizons.

There are two aspects of the optimal perturbation lifetime we wish to emphasise. Firstly, the long lifetime is not inherited from any underlying long-lived QNMs, such as those associated to conserved quantities in CFT. Rather, every QNM in the sum has a short lifetime, and it is only the non-normal effects that lead to the extended lifetime. Secondly, all evidence points to the lifetime of optimal perturbations being arbitrarily long but still \emph{finite}. While the lifetime of an optimal perturbation grows with the number of QNMs used in the sum, radial derivatives of an infinite sum fail to converge at the future horizon. Thus, our results concern only finite sums of QNMs which ultimately decay at asymptotically late times via QNM decay rates.

The optimal perturbations maximise the energy outside the horizon, corresponding to the choice of energy-norm when computing $G_W(\tau)$ \eqref{eq:energygrowth_subspace_W}. There are other choices of observables which can be computed with other choices of norms. An example is the $L^2([0,1])$ $\mu$-weighted norm $\langle \phi , \phi \rangle_{L^2} = \int_{0}^{1} dz \mu \overline{\phi} \phi$. This observable exhibits unbounded growth since it is insensitive to the value of $\p_{\tau}\phi$ -- one can simply choose initial data with an arbitrarily large $\p_{\tau}\phi$. On a related note, it would be interesting to investigate the potential physical consequences of  proposed orthogonality relations for QNMs \cite{Green:2022htq, London:2023aeo} (in the Kerr spacetime).

It is tempting to try to explain the existence of the optimal perturbations in this work by performing a `QNM decomposition' of classes of initial data that are supported near the horizon. Indeed, in the AdS case, the retarded Green's function appears to be meromorphic in $\omega$ with poles at QNM frequencies, and thus one may expect to be able to capture such initial data as a sum of QNMs. However, QNMs do not form a complete basis. For example, one can arrange initial data of compact support such that $\sup_\Sigma|\Phi|\neq 0$ for an arbitrarily long time, but with a vanishing QNM expansion -- see \cite{Warnick:2013hba, Warnick:2022hnc} for examples.\footnote{Note that here, in contrast, our perturbations are both arbitrarily long-lived (in terms of the energy outside the horizon) and explicitly constructed from finite sums of QNMs.}

The observable consequences of these perturbations is an important open question that, based on the uniformity of our results in different contexts, spans all domains where horizons are an integral part of the phenomenology. This includes cosmology, gravitational wave physics, analogue black holes \cite{Barcelo:2005fc}, and strongly-coupled many-body systems through AdS/CFT. In the gravitational wave context, a natural next step is to assess potential branch-cut contributions, not included in this work, as well as to construct optimal spin-2 perturbations of Kerr. Whether or not our long-lived perturbations can be seen in the wild likely depends on if they correspond to fine-tuned choices of initial data, in some suitable sense. Note that while the QNM coefficients are large and exhibit precise cancellations, the coefficients in the orthonormal basis for $M$ QNMs are not (see figure \ref{fig:dS_occupation}). Moreover, as long-lived phenomena, their role in an ensemble of initial states may be amplified. We would like to understand to what extent generic initial data contain such transients (e.g. binary black hole mergers), starting with the question of whether nearly-optimal initial data behave similarly.

Through AdS/CFT our results have implications for the thermalisation timescales of strongly coupled plasmas \cite{Balasubramanian:2011ur}, with applicability to the quark-gluon plasma \cite{Berges:2020fwq}. We have shown that there are initial states that, in a probe approximation, take an arbitrarily long time to thermalise. In other words, our work suggests that QNM decay rates may not be the ultimate arbiters of thermalisation timescales, due to non-normality. This is demonstrated through the time for QNM decay to affect CFT one-point functions (figure \ref{fig:SAdS_onepoint_abs_142856}) for such specially prepared states. Of course, whether or not the thermalisation timescale is truly arbitrary requires incorporating backreaction; the longer-lived states have a higher initial energy density near the horizon and likely backreact to a greater extent. It would also be interesting to assess the behaviour of higher-point functions. The role of the spatial distribution of initial conditions on the thermalisation timescale in CFT was investigated in earlier work \cite{Heller:2013oxa}.

There are also potential theoretical connections to other aspects of black hole physics, such as the Aretakis instability \cite{Aretakis:2012ei} and the firewall proposal \cite{Almheiri:2012rt,Almheiri:2013hfa}. In the latter, an infalling observer encounters high-energy quanta within a Planck length of the horizon. Here our perturbations can be localised arbitrarily close to the horizon, at least within the classical probe approximation. For example, in the de Sitter case, optimal perturbations are localised at a proper distance $ c\,H^{-1}M^{-1} + O(M)^{-2}$ from the horizon on a fixed $\tau$ slice. Thus the number of QNMs required for a Planck-sized transient is the number of Planck lengths that fit inside the Hubble radius -- on the order of $10^{61}$ QNMs in our late universe.

Finally, the transients we identified here have lifetimes which scale as $\tau \sim \log{M}$ where $M$ is the dimension of the Hilbert space, and thus are reminiscent of black hole scrambling time \cite{Hayden:2007cs, Sekino:2008he, Susskind:2011ap, Lashkari:2011yi}. It would be interesting to investigate this connection further.

\section*{Acknowledgements}
It is a pleasure to thank Daniel Are\'an, Oscar Dias, Michal Heller, Rodrigo Panosso Macedo, Christiana Pantelidou, Alexandre Serantes, Kostas Skenderis and David Turton for discussions. 
We thank Laura Sberna and Jan Wiersig for useful correspondence.
In addition BW would like to thank Michal Heller and Ghent University for hospitality.
The authors acknowledge the use of the IRIDIS High Performance Computing Facility, and associated support services at the University of Southampton, in the completion of this work.
JC is supported by the Royal Society Research Grant RF/ERE/210267.
BW is supported by a Royal Society University Research Fellowship and in part by the Science and Technology Facilities Council (Consolidated Grant ``Exploring the Limits of the Standard Model and Beyond'').

\appendix
\renewcommand{\addcontentsline}[3]{}

\section{Non-normality of $\mathcal{H}$}\label{sec:non_normality}
The Hamiltonian $\mathcal{H}$ \eqref{eq:block_hamiltonian} is not self-adjoint (Hermitian) with respect to the energy inner product \eqref{eq:energy_innerprod}, i.e. there exist $\xi_1, \xi_2$ such that
\be
\left<  \xi_1, \mathcal{H}^{\dagger}\xi_2 \right> \neq \left< \xi_1, \mathcal{H} \,\xi_2\right>, \label{HNSA}
\ee
where $\mathcal{H}^{\dagger}$ is the adjoint operator, defined as
\be
\left< \mathcal{H}^{\dagger} \xi_1, \xi_2 \right> = \left< \xi_1, \mathcal{H} \,\xi_2\right>, \qquad \forall \xi_1, \xi_2.
\ee
Demonstrating \eqref{HNSA} in generality is straightforward through integration by parts. In the dS$_{d+1}$ and Schwarzschild-AdS$_{d+1}$ case one picks up a boundary term at the future horizon,
\be
\left<  \xi_1, \mathcal{H}\xi_2 \right> - \left<\xi_1, \mathcal{H}^{\dagger}\xi_2\right> = -i\partial_\tau\bar{\phi}_1\partial_\tau\phi_2\Big|_{z=1},
\ee
while in Schwarzschild there is also a contribution from a flux at future null infinity, 
\be
\left<  \xi_1, \mathcal{H}\xi_2 \right> - \left<\xi_1, \mathcal{H}^{\dagger}\xi_2\right> = -i\left(\partial_\tau\bar{\phi}_1\partial_\tau\phi_2\Big|_{z=1} + \partial_\tau\bar{\phi}_1\partial_\tau\phi_2\Big|_{z=0}\right).
\ee
In the case $\xi_1 = \xi_2$ the right hand side is simply, $-i\mathcal{F}$, where $\mathcal{F}$ is the flux leaving the region \eqref{eq:FluxCC} and \eqref{eq:FluxFlat}. Another way to get this result is by applying time derivatives to the inner product: $\left<\xi_1, \mathcal{H}\xi_2 \right> - \left< \xi_1,\mathcal{H}^\dagger\xi_2 \right> = \left<\xi_1, i\partial_\tau\xi_2 \right> - \left<i\partial_\tau \xi_1, \xi_2 \right> = i\partial_\tau \left<\xi_1, \xi_2 \right>$, making the connection to the flux manifest. Note that taking further time derivatives generates higher-order operators which also evaluate to pure boundary terms, i.e.
\be
\left<\xi_1, \,:\mathrel{(\mathcal{H}-\mathcal{H}^\dagger)^n}:\,\xi_2 \right> = (i\partial_\tau)^{n-1}\left<\xi_1, \,(\mathcal{H}-\mathcal{H}^\dagger)\,\xi_2 \right>,
\ee
where $::$ denotes ordering such that $\mathcal{H}^\dagger$ is to the left of $\mathcal{H}$.

Normal operators are characterised by $\left[\mathcal{H}, \mathcal{H}^\dagger\right] = 0$, or equivalently, by having a complete orthonormal set of eigenfunctions. Thus while Hermiticity implies normality, non-Hermiticity does not imply non-normality, because a non-Hermitian operator may still commute with its adjoint. For example, a unitary operator is a non-Hermitian normal operator. 

To demonstrate non-normality, we restrict to the subspace of perturbations $W$, formed from a sum of $M = \dim(W)$ QNMs, as introduced in section \ref{sec:energy_growth}. In the $W$ subspace all perturbations take the form
\be
\xi = \sum_{n=1}^{M}c_n\tilde{\xi}_n = \sum_{n=1}^{M} d_n \psi_n ,
\ee
with $\vec{d} = U_W\vec{c}$ determined by Gram-Schmidt. 
Without loss of generality we can thus express all perturbations according to their $\vec{d}$ vector in this decomposition. In this basis the energy inner product becomes simply
\be
\left<\xi_1, \xi_2\right> = \left<\vec{d}_1, \vec{d}_2\right>_2 \equiv \vec{d}^{\,*}_1\, \vec{d}_2.
\ee
In particular, we can write the Hamiltonian $\mathcal{H}_W$ in this basis as the $M\times M$ matrix,
\be
H_W = U_W D_W U_W^{-1}.
\ee
Thus its Hermitian adjoint, $H_W^\dagger$, defined as
\be
\left<H_W^\dagger\vec{d}_1, \vec{d}_2\right>_2 = \left<\vec{d}_1, H_W\vec{d}_2\right>_2,\qquad \forall \vec{d}_1, \vec{d}_2
\ee
is the conjugate transpose of $H_W$, 
\be
H_W^\dagger = (U_W D_W U_W^{-1})^*.
\ee

Non-normality of the system, when restricted to the subspace of perturbations $W$, can then be assessed by computing the matrix norm $\|[H_W^\dagger, H_W]\|_2^2$. For example, if the eigenfunctions $\{\tilde{\xi}_n\}$ were unitarily related to the orthonormal basis $\{\psi_n\}$ (i.e. $U_W$ a unitary matrix) then this commutator would vanish.
We have computed this commutator for the cases considered in this paper, as detailed in table \ref{tab:nonnormality}, clearly demonstrating their non-normality.
\begin{table}[h!]
\begin{center}
\begin{tabular}{ |c|c|c|c|c| } 
 \hline
 Spacetime & Perturbation & $\dim(W)$ & Section & $\|[H_W^\dagger, H_W]\|_2^2$\\ 
\hline
 dS$_4$ & $m^2 = 2$, $l=0$ & 16 & \ref{sec:dS_energygrowth} & $4.5\times 10^8$ \\ 
 dS$_4$ & $m^2 = 2$, $l=0$ & 32 & \ref{sec:dS_energygrowth} & $1.0\times 10^{11}$\\ 
 dS$_4$ & $m^2 = 2$, $l=0$ & 64 & \ref{sec:dS_energygrowth} & $2.5\times 10^{13}$ \\ 
 dS$_4$ & $m^2 = 50$, $l=0$ & 32 & \ref{sec:dS_energygrowth} & $1.0\times 10^{11}$\\ 
 Schw.-AdS$_4$ & $\Delta = 2$, $\vec{k}=0$ & 14 & \ref{sec:ads_numerics} & $2.4\times 10^8$ \\ 
 Schw.-AdS$_4$ & $\Delta = 2$, $\vec{k}=0$ & 28 & \ref{sec:ads_numerics} & $5.6\times 10^{10}$ \\ 
 Schw.-AdS$_4$ & $\Delta = 2$, $\vec{k}=0$ & 56 & \ref{sec:ads_numerics} & $1.4\times 10^{13}$ \\
  Schw. ($3+1$, asy. flat) & $s = 2$, $l = 2$ & 10 & \ref{flat_section} & $1.7\times 10^3$\\
   Schw. ($3+1$, asy. flat) & $s = 2$, $l = 2$ & 19 & \ref{flat_section} & $4.7\times 10^5$\\
 Schw. ($3+1$, asy. flat) & $s = 2$, $l = 2$ & 39 & \ref{flat_section} & $2.4\times 10^8$\\
 \hline
\end{tabular}
\end{center}
\caption{Demonstration of non-normality of the subspace Hamiltonian $H_W$ for the various spacetime and perturbation examples studied in this paper. A normal Hamiltonian would satisfy $[H_W^\dagger, H_W] = 0$.}
\label{tab:nonnormality}
\end{table}

\section{Proof of \eqref{HWdef} \label{sec:HWproof}}

Using \eqref{eq:xiexpansion_M_orthbasis}, the change of basis $\tilde{\xi}_n = \sum_{m=1}^{M} (U_W)_{mn} \psi_m$, orthonormality $\langle \psi_j , \psi_k \rangle = \delta_{jk}$, and $\vec{d}=U_W\vec{c}$, one has

\bea
\|\mathcal{H}_W \xi(0,z)\|_{E}^{2} &=& \langle \mathcal{H}_W \xi(0,z) , \mathcal{H}_W \xi(0,z) \rangle \\
&=& \sum_{n,m=1}^{M}c_{n}^*c_m \langle \mathcal{H}_W \tilde{\xi}_n , \mathcal{H}_W \tilde{\xi}_m \rangle \\
&=& \sum_{n,m=1}^{M}c_{n}^*c_m \omega_{n}^*\omega_m \langle \tilde{\xi}_n , \tilde{\xi}_m \rangle \\
&=& \sum_{n,m=1}^{M}\sum_{j,k=1}^{M}(\omega_n c_n)^* (\omega_m c_m) ((U_W)_{jn})^* (U_W)_{km} \langle \psi_j, \psi_k \rangle \\
&=& \sum_{n,m=1}^{M}\sum_{j=1}^{M} ((U_W)_{jn} \omega_n c_n)^* \((U_W)_{jm} \omega_m c_m \) \\
&=& \(U_W D_W \vec{c}\)^*\(U_W D_W \vec{c}\) \\
&=& \|U_W D_W \Vec{c}\,\|_{2}^{2} \\
&=& \|U_W D_W U_{W}^{-1} \,\Vec{d}\,\|_{2}^{2}.
\eea
Similarly,
\bea
\|\xi(0,z)\|_{E}^2 &=& \langle \xi(0,z) , \xi(0,z) \rangle \\
&=& \sum_{n,m=1}^{M}c_{n}^*c_m \langle \tilde{\xi}_n , \tilde{\xi}_m \rangle \\
&=& \sum_{n,m=1}^{M}\sum_{j,k=1}^{M}c_n^* c_m ((U_W)_{jn})^* (U_W)_{km} \langle \psi_j, \psi_k \rangle \\
&=& \(U_W \vec{c}\)^*\(U_W \vec{c}\) \\
&=& \|U_W \Vec{c}\,\|_{2}^{2} \\
&=& \|\Vec{d}\,\|_{2}^{2},
\eea
such that
\be
\|\mathcal{H}_W\|_{E}^{2}=\sup_{\xi(0,z) \in W} \frac{\|\mathcal{H}_W \xi(0,z)\|_{E}^{2}}{\|\xi(0,z)\|_{E}^{2}} = \max_{\vec{d}\in \mathbb{C}^M} \frac{\|U_W D_W U_{W}^{-1} \,\Vec{d}\,\|_{2}^{2}}{\|\Vec{d}\,\|_{2}^{2}} = \|U_WD_WU_{W}^{-1}\|_{2}^2.
\ee

\section{Chebyshev spectral methods}\label{sec:spectral_methods}
Spectral methods for finite, non-periodic domains approximate functions by truncated expansions of Chebyshev polynomials of the first kind,
\be
T_n(x) = \cos(n \arccos(x)), \qquad -1 \leq x \leq 1,
\ee
up to some order $N$ as follows,
\be
\label{eq:cheb_interpolant}
f(z) \approx f_N(z) \equiv \sum_{k=0}^{N}\tilde{c}_k \tilde{T}_k(z),
\ee
where $\tilde{T}_k(z) \equiv T_k(\frac{2z-a-b}{b-a})$, i.e. the argument of the polynomials is rescaled to accommodate $z\in [a,b]$. Note that in this work we have $z\in[0,1]$ for both dS$_{d+1}$ and Schwarzschild-AdS$_{d+1}$, hence $a=0$ and $b=1$ for our purposes. 

The interpolant function, $f_N(z)$, is uniquely determined by choosing a grid of $N+1$ points, here the (shifted) Chebyshev-Lobatto points 
\be
\label{eq:cheb_grid}
z_j=\frac{a+b}{2}-\frac{b-a}{2}\cos\(\frac{j}{N}\pi\),\qquad j=0,1,\ldots,N
\ee
such that $f(z_j)=f_N(z_j)$. The coefficients $\tilde{c}_k$ are determined using the orthogonality relations \cite{abramowitz+stegun}
\be
\sum_{k=0}^N (2-\delta_{k0}-\delta_{kN})\tilde{T}_i(z_k)\tilde{T}_j(z_k) = (1+\delta_{i0}+\delta_{iN})N \delta_{ij}
\ee
in \eqref{eq:cheb_interpolant}, resulting in,
\be
\label{eq:cheb_coeffs}
\tilde{c}_k=\frac{1}{N(1+\delta_{k0}+\delta_{kN})}\sum_{j=0}^{N}(2-\delta_{j0}-\delta_{jN})\tilde{T}_k(z_j)f(z_j).
\ee
Once we have determined $f_N(z)$ in \eqref{eq:cheb_interpolant} with the coefficients \eqref{eq:cheb_coeffs}, the approximation of the $m^{\text{th}}$ derivative of $f(z)$ at the grid points can be stated as
\be
f^{(m)}(z_j)\approx f^{(m)}_N(z_j)=\sum_{k=0}^N(D_{N}^{(m)})_{jk}f(z_k),
\ee
where $D_{N}^{(m)}$ is the $(N+1) \times (N+1)$ $m^{\text{th}}$-order Chebyshev differentiation matrix.

In this way, evaluating the respective functions on the grid points and substituting radial derivatives with differentiation matrices of the appropriate order, the Hamiltonian in \eqref{eq:block_hamiltonian} is discretised into a $2(N+1) \times 2(N+1)$ matrix, $H$, and its eigenfunctions, $\xi_n$, into $2(N+1)$ eigenvectors $\vec{\xi}_n$ (at $\tau=0$).

Similarly, we can evaluate the energy inner product \eqref{eq:energy_innerprod} for functions in the Chebyshev approximation. The main ingredient we need is,
\be
\label{eq:cheb_integral}
\int_a^b dz\,f(z) \approx \int_a^b dz\,f_N(z) = (b-a)\sum_{k=0}^{\lfloor N/2 \rfloor}\frac{\tilde{c}_{2k}}{1-4 k^2}.
\ee
Using \eqref{eq:cheb_integral} and \eqref{eq:cheb_coeffs}, the terms appearing in the energy inner product \eqref{eq:energy_innerprod} can be approximated as
\be
\label{eq:cheb_innerprod_genterm}
\int_a^b dz\,\mu(z) f(z)g(z) \approx \int_a^b dz\,(\mu fg)_N(z) = \vec{f}^{\,\,T}C_N[\mu]\,\vec{g},
\ee
where $\vec{f} = (f(z_0) \;f(z_1) \ldots f(z_N))^T$, and $C_N[\mu]$ is the $(N+1)\times(N+1)$ $\mu$-weighted quadrature diagonal matrix with elements
\be
\label{eq:cheb_quadmatrix}
(C_N[\mu])_{ij}=\delta_{ij}\frac{(b-a)}{N}(2-\delta_{i0}-\delta_{iN})\mu(z_i)\sum_{k=0}^{\lfloor N/2 \rfloor} \frac{\tilde{T}_{2k}(z_i)}{(1+\delta_{2k,0}+\delta_{2k,N})(1-4k^2)}.
\ee
Following this procedure, \eqref{eq:energy_innerprod} is discretised into 
\be
\label{eq:cheb_innerprod_discr}
\langle \vec{\xi}_1, \vec{\xi}_2 \rangle = \vec{\xi}_1^{\,*} G \,\vec{\xi}_2 = \vec{\xi}_1^{\,*} F^* F \,\vec{\xi}_2 = (F\vec{\xi}_1)^{*}\, F\vec{\xi}_2 ,
\ee
with $G$ the $2(N+1)\times2(N+1)$ matrix defined as
\be
G= \frac{1}{2} \begin{pNiceArray}{c|c}
   (D_N^{(1)})^TC_N[p]\, D_N^{(1)} + C_N[q] & \Block{1-1}<\Large>{0} \\
  \hline
  \Block{1-1}<\Large>{0} & C_N[w]
\end{pNiceArray},
\ee
for the $w(z)$, $p(z)$, $q(z)$ in each spacetime. In practice, in order to minimise the loss of accuracy resulting from approximating the integrand in \eqref{eq:cheb_innerprod_genterm} as a single Chebyshev expansion (even if we need to evaluate both functions on the grid, $\vec{f}$ and $\vec{g}$), we first compute the $G$ matrix in a grid of double resolution $\tilde{N} = 2 N + 1$. The interpolation $(\tilde{N}+1)\times(N+1)$ matrix, $\mathcal{I}$, connecting both grids is obtained using \eqref{eq:cheb_coeffs} and demanding that
\be
f_{\tilde{N}}(\tilde{z}_i)=\sum_{j=0}^N (\mathcal{I})_{ij} f_N(z_j),
\ee
where $\tilde{z}_i$ ($i=0,1,\ldots,\tilde{N}$) are the Chebyshev-Lobatto points in the new grid. Thus we have
\be
(\mathcal{I})_{ij} = \frac{(2-\delta_{j0}-\delta_{jN})}{N}\sum_{k=0}^N\frac{\tilde{T}_k(\tilde{z}_i)\tilde{T}_k(z_j)}{(1+\delta_{k0}+\delta_{kN})},
\ee
and the $G$ matrix we use in \eqref{eq:cheb_innerprod_discr} is given by the interpolation back to the original grid
\be
G=\frac{1}{2} \begin{pNiceArray}{c|c}
   \mathcal{I}^T & \Block{1-1}<\Large>{0} \\
  \hline
  \Block{1-1}<\Large>{0} & \mathcal{I}^T
\end{pNiceArray} \begin{pNiceArray}{c|c}
   (D_{\tilde{N}}^{(1)})^TC_{\tilde{N}}[p] D_{\tilde{N}}^{(1)} + C_{\tilde{N}}[q] & \Block{1-1}<\Large>{0} \\
  \hline
  \Block{1-1}<\Large>{0} & C_{\tilde{N}}[w]
\end{pNiceArray} \begin{pNiceArray}{c|c}
   \mathcal{I} & \Block{1-1}<\Large>{0} \\
  \hline
  \Block{1-1}<\Large>{0} & \mathcal{I}
\end{pNiceArray}.
\ee

One last comment is in order. The approximation to the integral in \eqref{eq:cheb_innerprod_genterm} is only valid when the weight function $\mu(z)$ does not diverge at any of the points in the grid, as can be seen from \eqref{eq:cheb_quadmatrix}. One could encounter, however, a situation where $\mu(z)$ blows up at some $z=z_j$ but the exact integral is still convergent. This is the case, for instance, of the static patch of dS$_2$ for $l=0$, where $w(z)=(2\sqrt{z})^{-1}$ and $q(z)=m^2 (2\sqrt{z})^{-1}$ in \eqref{eq:dS_w} and \eqref{eq:dS_q}. For divergences of this type, $\mu(z) \propto z^{-1/2}$, the idea is to use a better approximation in which we leave $\mu(z)$ outside the Chebyshev expansion, i.e.
\be
\int_a^b dz\,\mu(z) f(z)g(z) \approx \int_a^b dz\,\mu(z)(fg)_N(z) = \vec{f}^{\,\,T}\mathcal{C}_N[\mu]\,\vec{g},
\ee
where now we have
\be
(\mathcal{C}_N[\mu])_{ij}=\delta_{ij}\frac{(2-\delta_{i0}-\delta_{iN})}{N}\sum_{k=0}^{N} \frac{\tilde{T}_{k}(z_i)}{(1+\delta_{k0}+\delta_{kN})}I_k[\mu],
\ee
having defined
\be
\label{eq:cheb_specialcase}
I_k[\mu] \equiv \int_a^b dz\,\mu(z)\tilde{T}_k(z).
\ee
Given $\mu(z) = \alpha z^{-1/2}$ and $a=0$, $b=1$, concerning dS$_2$ at $l=0$, the integral \eqref{eq:cheb_specialcase} has a closed form expression
\be
I_k[\mu]= \alpha \frac{2}{1-4k^2}.
\ee

\bibliographystyle{JHEP}
\bibliography{refs}

\end{document}